\documentclass[11pt]{article}

\usepackage[final]{acl}
\usepackage{times}
\usepackage{latexsym}

\usepackage[table]{xcolor}
\usepackage{booktabs}
\usepackage{array}
\usepackage{soul}

\usepackage[LGR,T1]{fontenc}
\DeclareFontFamilySubstitution{LGR}{ptm}{udidot}

\usepackage{textalpha}

\definecolor{headerbg}{HTML}{F3F4F6}      
\definecolor{factnum}{HTML}{5B7F4A}       

\definecolor{qrelbg}{HTML}{DCEBFA}        
\definecolor{qrelborder}{HTML}{A9C7E8}    

\definecolor{reasonbg}{HTML}{EAF7DA}      
\definecolor{conclusionbg}{HTML}{FFE2CC}  
\definecolor{baselinebg}{HTML}{FDE2E2}    

\definecolor{textgray}{HTML}{222222}

\newcommand{\best}[1]{\cellcolor{qrelbg}\textbf{#1}}

\newcommand{\localbest}[1]{\cellcolor{qrelbg}#1}

\newcommand{\lawhl}[1]{\begingroup\sethlcolor{qrelbg}\hl{#1}\endgroup}

\usepackage[utf8]{inputenc}

\usepackage{microtype}

\usepackage{inconsolata}

\usepackage{graphicx}

\title{GreekBarRetrieval: A Benchmark for Greek Statutory Retrieval}

\author{
Ernest Beta\textsuperscript{1}
\quad
Odysseas S. Chlapanis\textsuperscript{1,2}
\\[0.25em]
\textbf{Dimitrios Galanis\textsuperscript{2,3}
\quad
Ion Androutsopoulos\textsuperscript{1,2}}
\\[0.5em]
\textsuperscript{1}Department of Informatics, Athens University of Economics and Business, Greece
\\
\textsuperscript{2}Archimedes, Athena Research Center, Greece
\\
\textsuperscript{3}Institute for Language and Speech Processing, Athena Research Center, Greece
}

\begin{document}
\maketitle
\begin{abstract}
Statutory retrieval is necessary for citation-grounded legal question answering, but remains underexplored for Greek. We introduce GreekBarRetrieval, a public retrieval benchmark derived from, and complementing GreekBarBench \cite{chlapanis-etal-2025-greekbarbench}, which did not include retrieval. The new benchmark comprises 283 bar-exam questions, each accompanied by the facts of the case it refers to, and 6,308 candidate statutory articles to retrieve from. Questions and facts are stated in everyday language, but need to be mapped to the formal terminology of statutes and their abstract legal concepts. A further complication is that not all of the case facts are relevant to each question of a case. Experimenting with three BM25 variants and nine dense retrievers, we find that vanilla dense retrieval far outperforms vanilla sparse retrieval in Recall@100. However, LLM-based query reformulation helps \textsc{BM25} close that gap, while also improving dense retrieval. With a ten-round \textsc{ReAct}-like LLM reformulation loop that we introduce, \textsc{BM25} improves further in Recall@100 and obtains the best nDCG and MAP scores of all tested retrievers. Query reformulation also outperforms pseudo-relevance feedback, sparse-dense fusion, and English translation.
\end{abstract}

\begin{table}[!t]
\centering
\footnotesize
\setlength{\tabcolsep}{0pt}
\renewcommand{\arraystretch}{1.05}
\begin{tabular}{p{0.99\linewidth}}
\toprule
\rowcolor{headerbg}
\textbf{Facts} \\
\midrule
\vspace{-2mm}
{\color{factnum}\textbf{[1]}} A, a heart patient in crisis, goes to B's only overnight pharmacy for life-saving medication. \\
{\color{factnum}\textbf{[2]}} B refuses because of a personal dispute, despite C's urgent request and A's imminent collapse. \\
{\color{factnum}\textbf{[3]}} A says: ``Let me die, and let him bear the blame.'' \\
{\color{factnum}\textbf{[4]}} C tries to give A the medication; when B attempts to stop him, C strikes B and gives it to A. \\
\midrule
\rowcolor{headerbg}
\textbf{Question} \\
\midrule
Which criminal offenses were committed by B and C? \\
\midrule
\rowcolor{headerbg}
\textbf{Gold Relevant Articles} \\
\midrule
\fcolorbox{qrelborder}{qrelbg}{\strut \texttt{CrimC::42}}
\hspace{0.25em}
\fcolorbox{qrelborder}{qrelbg}{\strut \texttt{CrimC::299}}
\hspace{0.25em}
\fcolorbox{qrelborder}{qrelbg}{\strut \texttt{CrimC::15}}
\hspace{0.25em}
\fcolorbox{qrelborder}{qrelbg}{\strut \texttt{CrimC::22}} \\
\midrule
\rowcolor{headerbg}
\textbf{Official Solution (summarized for the example)} \\
\midrule
B refused life-saving medication despite A's imminent collapse and his special duty to act.
\textit{Attempted homicide by omission}.
\lawhl{Arts. 42, 299, 15 CrimC}. \\[0.25em]

C used force only to secure the medication and protect A's life.
\textit{Defense of a third person}.
\lawhl{Art. 22 CrimC}. \\
\bottomrule
\end{tabular}
\vspace{-2mm}
\caption{Example GreekBarRetrieval instance. The retrieval query is \textit{Question + Facts} and the gold statutory articles are those cited in the official solution. Articles are identified as \texttt{PREFIX::ARTICLE}.  \texttt{CrimC} is the Greek Penal Code.
Example translated to English.}
\vspace{-5mm}
\label{tab:gbb_example}
\end{table}

\section{Introduction}
Legal question answering should ground its answers in retrieved relevant authorities, such as applicable statutory articles. Interpretability is a central requirement in the legal domain~\citep{martinezgil2023survey}, while precise retrieval supports citations and enables users to verify model claims~\citep{pipitone2024legalbenchrag}. Retrieval also affects answer quality; prior work shows that providing relevant legal passages can substantially improve downstream legal question answering~\citep{zheng2025reasoningfocused}. Legal retrieval is complicated, however, by a vocabulary mismatch; questions often describe events in everyday language, whereas the applicable authorities express the governing rules through specialized terminology and abstract legal concepts. In Greek, rich morphology creates additional surface variation, making exact term matching less reliable for sparse retrievers such as BM25~\citep{ntais2006development,papantoniou2024nlp}. Dense retrievers can bridge some of this mismatch, but may underweight exact lexical cues (e.g., required exact statutory terms, article references), and highly rank articles that resemble the case facts without addressing the question-specific legal issue.

Existing legal retrieval benchmarks cover statutes, case law, contracts, and legal question answering~\citep{pipitone2024legalbenchrag,zheng2025reasoningfocused,louis2021bsard,su-etal-2024-stard,ma2021lecard,li2024lecardv2,joshi-etal-2024-il,goebel2026coliee,mahari2024lepard}. We introduce \textbf{GreekBarRetrieval}, a public benchmark derived from GreekBarBench~\citep{chlapanis-etal-2025-greekbarbench}, which used questions from the Greek Bar exams. GreekBarBench evaluates LLMs on legal questions from the Greek Bar exams without involving true retrieval; instead, each question is accompanied by the gold relevant articles and distractors (irrelevant articles). By contrast, GreekBarRetrieval contains 283 Greek bar examination questions, each accompanied by case facts (both from GreekBarBench), and a pool of 6,308 candidate statutory articles. For each question, the relevant statutes need to be retrieved from the pool. The retrieval query consists of the question and the given facts. The gold statutes are obtained from the official solution (Table~\ref{tab:gbb_example}). Unlike GreekBarBench, which evaluates complete answers with citations, but without true retrieval, GreekBarRetrieval focuses on statute retrieval only. Hence, it complements GreekBarBench by evaluating the retrieval abilities of legal Retrieval Augmented Generation (RAG) LLMs. 
To our knowledge, GreekBarRetrieval is the first retrieval benchmark for Greek statutory law. For experimental purposes, we also release GreekBarRetrieval in a machine-translated English form. 

To demonstrate the usage of GreekBarRetrieval and establish baselines, 
we use it to compare BM25 (with three pre-processing variants  to cope with Greek morphology) and nine dense embedding models. Vanilla dense retrieval far outperforms vanilla sparse retrieval; \texttt{Gemini-001} reaches 0.77 Recall@100 against 0.36 for \textsc{BM25}.
However, LLM-based query reformulation largely closes that gap; vanilla reformulation improves \textsc{BM25} to 0.60 Recall@100, while also improving dense retrieval. Furthermore, with a ten-round \textsc{ReAct}-like \cite{yao2023react} reformulation loop that we introduce, \textsc{BM25} obtains 0.67 Recall@100 and the highest nDCG and MAP scores of all methods tested.
We also show that query reformulation has a much more substantial effect on BM25 than tuning its parameters or varying pre-processing, and that it outperforms pseudo-relevance feedback in sparse and dense retrieval, as well as sparse-dense fusion.

Our main contributions are: (1) We introduce \textbf{GreekBarRetrieval},\footnote{%
\url{https://huggingface.co/datasets/AUEB-NLP/greek-bar-bench};
code at \url{https://github.com/nlpaueb/greek-bar-bench}.
}
a public benchmark for Greek statutory retrieval, based on bar-exam questions.
The new benchmark is derived from, and complements GreekBarBench, which did not include
true retrieval.
(2) We provide \textbf{experimental results} for BM25 (with three Greek-specific pre-processors) and 9 dense  retrievers, also considering LLM query reformulation, pseudo-relevance feedback, sparse-dense fusion, and translation to English. We show that LLM \textbf{query reformulation} is particularly beneficial for BM25, while also improving dense retrieval.
(3) We introduce a \textbf{\textsc{ReAct}-like LLM query reformulation loop}, which vastly improves the performance of BM25, helping it obtain the highest nDCG and MAP scores among all methods tested, including dense retrievers.


\section{GreekBarRetrieval}
\label{sec:datasets}

\subsection{Derivation from GreekBarBench}
\label{subsec:derivation}

GreekBarRetrieval is derived from GreekBarBench, a benchmark based on Greek Bar examinations across five legal areas: Civil (Αστικό), Criminal (Ποινικό), Commercial (Εμπορικό), Public Law (Δημόσιο), and the Lawyers' Code (Κώδικας Δικηγόρων)~\citep{chlapanis-etal-2025-greekbarbench}. Each GreekBarBench instance contains the facts of a case, a legal question about the case, a mix of relevant and irrelevant statutory articles supplied as context, and an official solution. We keep the questions, facts and the articles cited by the solution, and discard the supplied context; the cited articles become the ground truth (gold articles) for retrieval.

\subsection{Benchmark Construction}
\label{subsec:construction}
We construct the \textbf{retrieval pool} (the set to retrieve articles from) from the statutory articles of \emph{all} the Greek legal code documents. Each article is identified by its legal source and article number, such as \texttt{CivC::11}, \texttt{CivProc::41}, or \texttt{CrimC::42}. Paragraph and subsection numbers are not included, because retrieval is evaluated at the level of complete articles. This process produces 6,308 candidate articles from 23 legal sources (prefixes).

For each of the 310 GreekBarBench questions, we convert the articles cited in the official
solution into the same \texttt{SOURCE::ARTICLE} identifiers, removing repetitions. We exclude
27 questions, 16 whose solution cites no articles and 11 citing at least one article absent
from the retrieval pool. This leaves 283 queries: 257 publicly released and, following
\textsc{GreekBarBench}, 26 from the 2024 exams held out as a semi-private test set.


\subsection{Dataset Statistics}
\label{subsec:statistics}

As already explained, GreekBarRetrieval contains 283 queries and 6,308 candidate articles (retrieval pool) from 23 legal sources (prefixes). Table~\ref{tab:gbb_length_stats} reports length statistics. Queries concatenate a legal question with the facts of a case, so the task is to retrieve short statutory articles from much longer descriptions (queries) of legal cases.

\begin{table}[t]
\centering
\small
\setlength{\tabcolsep}{4pt}
\begin{tabular}{lrrrrr}
\toprule
\textbf{Unit} & \textbf{Count} & \textbf{Mean} & \textbf{Median} &
\textbf{P95} & \textbf{Max} \\
\midrule
articles & 6{,}308 & 108.04 & 58  & 372 & 3{,}778 \\
queries  & 283     & 290.33 & 219 & 773 & 1{,}013 \\
\bottomrule
\end{tabular}
\caption{Text length \textbf{statistics} for GreekBarRetrieval. Count is the number of articles or queries. Mean, Median, P95, and Max are the mean, median, 95th percentile, and maximum lengths in words, respectively.}
\vspace{-6mm}
\label{tab:gbb_length_stats}
\end{table}

GreekBarRetrieval contains 775 \textbf{relevance judgments} (total gold article citations, for all queries) covering 465 distinct articles, or 7.37\% of the retrieval pool. Each query has 2.74 gold articles on average, with a median of 2 and a maximum of 27. Overall, 56.9\% of queries have more than one relevant article, 9.9\% have more than five, and 21.2\% cite articles from more than one legal source.

We also retain the five legal areas assigned to sources (prefixes) in GreekBarBench~\citep{chlapanis-etal-2025-greekbarbench} for analysis purposes. A source may be associated with more than one area. Appendix~\ref{app:legal_area_statistics} reports statistics by legal area, while Appendix~\ref{app:prefix_area_mapping} provides the mapping from prefixes to areas.

\subsection{Retrieval Task}
\label{subsec:task}

The GreekBarRetrieval task is particularly challenging for three reasons. First, the questions and (especially) the facts are stated in everyday language (Table~\ref{tab:gbb_example}), so the retrieval task requires mapping questions and facts
to statutory terminology and abstract legal concepts. 
Second, 
the questions come from bar exam files. Each file provides the facts of a particular case, then asks multiple questions about the case. Some of the facts of a case may be irrelevant to some of the questions of the case.   
Consequently, a query (which concatenates a question with all the facts of its case) may contain details irrelevant to the particular legal issue being asked. 
Third, 
questions often require combining information distributed across multiple articles from different sources; 56.9\% of queries have multiple gold relevant articles, and 21.2\% have gold articles from multiple legal sources (prefixes). 

\section{Experimental Setup and Retrievers}
\label{sec:experiments}

\subsection{Evaluation Metrics}
\label{subsec:evaluation_metrics}

We report standard retrieval metrics  \cite{manning2008introduction}: nDCG@10, nDCG@100, and MAP@100 for ranking quality; Recall@10 and Recall@100 for evidence coverage.
Coverage matters because legal questions often require
several articles, not just one.
Recall@100 is our primary coverage metric and, intuitively, it assesses the
extent to which the required articles are present in the context of the LLM
that generates the answer, assuming the top 100 retrieved articles are all
passed to the generator.\footnote{If re-rankers are involved, they are  typically applied to the top retrieved articles, hence Recall@\(k\) can be seen as checking if the required documents will be available to the re-ranker.}  
Recall@10 reflects coverage when the generator LLM has a smaller context available. We compute all metrics using the \texttt{pytrec\_eval} module provided by the \texttt{pytrec-eval-terrier} package.\footnote{\url{https://pypi.org/project/pytrec-eval-terrier}}


\subsection{Vanilla Dense and Sparse Retrievers}
\label{subsec:retrieval_baselines}

We evaluate three sparse BM25 baselines that differ in their pre-processing of Greek text: \textsc{BM25-GreekStemmer}, \textsc{BM25-spaCy}, and \textsc{BM25-gr-nlp-toolkit} (Appendix~\ref{app:retriever_details}). We also evaluate nine dense embedding models that vary in size and language coverage,   including both general-purpose and legal-domain models, served either locally or through an API: \texttt{Gemini-001}, \texttt{Qwen3-8B}, \texttt{Euler-Legal-V1}, \texttt{Qwen3-4B}, \texttt{Jina-v5-small}, \texttt{Arctic-v2}, \texttt{Qwen3-0.6B}, \texttt{EmbGemma-300M}, and \texttt{Nomic-v1.5}. Full model identifiers and implementation details are provided in Appendix~\ref{app:retriever_details}. \textsc{BM25-GreekStemmer} and \textsc{BM25-spaCy} are the two best performing BM25 variants and are effectively indistinguishable; we use \textsc{BM25-GreekStemmer} as the representative BM25 variant in the main experimental results below.

\subsection{BM25 Parameter Tuning}
\label{sec:bm25_parameter_tuning}

As a sensitivity analysis, we also sweep the values of $k_1$ and $b$ in \textsc{BM25-GreekStemmer} using the original and LLM-rewritten queries. This sweep is diagnostic only; it tests whether the sparse baseline is unusually sensitive to parameter values other than the default ones. Because the gains are small (Table~\ref{tab:interventions}) all other BM25 experiments reported here use the default $k_1$ and $b$ values (Table~\ref{tab:app_sparse_retrievers}).

\subsection{English Translation Baselines}
\label{subsec:english_translation_baseline}

For additional vanilla baselines, we translate GreekBarRetrieval into English to test whether sparse and dense retrievers improve when both the queries and candidate articles are in English. We translate questions, facts, and articles using \texttt{openai/gpt-oss-120b}~\citep{openai2025gptoss120bgptoss20bmodel}, while preserving query identifiers, document identifiers, and relevance judgments. The translation prompt is shown in Figure~\ref{fig:en_translation_prompt} (Appendix~\ref{app:translation_reformulation}). We evaluate only the vanilla sparse and dense retrievers (\S\ref{subsec:retrieval_baselines}) on the translated benchmark. These runs use the same setup as the original Greek baselines, with default English rather than Greek-specific pre-processing for sparse retrieval (Table~\ref{tab:app_sparse_retrievers}).

\subsection{Pseudo-Relevance Feedback (PRF)}
\label{subsec:pseudo_relevance_feedback}

To test if retrieved articles can improve the query without using LLM-based query reformulation, we apply pseudo-relevance feedback (PRF). For sparse retrieval, we use a simple TF--IDF term-expansion PRF heuristic inspired by classical local-analysis methods~\citep{xu1996query}. Specifically, we initially use BM25 to retrieve the top 10 articles. From this set, we select the 30 terms with the highest TF-IDF scores and append them to the original query before running BM25 again. We keep the retrieved articles (and ranking) of the second run.

For dense retrieval, we apply positive-feedback Rocchio in the embedding space~\citep{rocchio1971relevance}. After an initial dense retrieval pass, we take the top 10 retrieved articles as pseudo-relevant, compute the centroid $c$ of their article embeddings, and form a new query vector
$q_{\mathrm{PRF}}=\mathrm{norm}(\alpha q+\beta c)$,
where $q$ is the original query embedding and $\mathrm{norm}$ denotes L2 normalization. We use $\alpha=1.0$ and $\beta=0.25$, which give the original query more weight in order to reduce query drift. 
The first-pass ranking is used only to construct the feedback centroid; for evaluation, we retain the retrieved articles (and ranking) of the second pass (with $q_{\mathrm{PRF}}$).

\subsection{Sparse-Dense Fusion (RRF)}  
\label{subsec:sparse_dense_fusion}

To test whether sparse and dense retrievers recover complementary evidence, we combine \textsc{BM25-GreekStemmer} in turn with each one of the eight locally hosted dense retrievers
(Table ~\ref{tab:app_dense_retrievers}), i.e., excluding only \texttt{Gemini-001}. For each pair of \textsc{BM25-GreekStemmer} and dense retriever, we fuse their top 100 results using Reciprocal Rank Fusion (RRF)~\citep{cormack2009rrf} with default $k$ value ($k=60$). Since RRF uses rank positions rather than retrieval scores, it does not require score calibration between BM25 and the dense models. 

\subsection{LLM Query Reformulation} 
\label{subsec:llm_query_reformulation}

As discussed in \S\ref{subsec:task}, the questions and (especially) facts of GreekBarRetrieval are stated in everyday language, whereas the articles to be retrieved use specialized legal terminology and abstract legal concepts. Furthermore, each  retrieval query concatenates a question with all the facts of its case, including possibly irrelevant facts. We, therefore, use an LLM to reformulate the queries, hoping to (a) keep only facts related to the legal issue of the question, and (b) bring the queries closer to the legal terminology and concepts of the authorities.  


A different reformulation strategy is used for sparse and dense retrieval.
For sparse retrieval, the LLM uses the question to identify the relevant parts of the facts and express the legal issue of the question in legal terminology. It produces a compact list of keywords (and short phrases), omitting secondary narrative details,  while preserving any explicit references to articles, laws, or decisions.
For dense retrieval, the LLM follows the same principle, but produces a concise rewritten query in the form of a prose statement, instead of a keyword list. The rewrite aims to preserve the question and the facts needed to express the legal issue, using Greek legal terminology, and removing secondary details that may distract the embedding model. All query reformulations are generated with \texttt{gpt-oss-120b}. The full prompts are shown in Figures~\ref{fig:sparse_rewrite_prompt} and~\ref{fig:dense_rewrite_prompt}.

\subsection{ReAct-BM25}
\label{subsec:agentic_bm25_retrieval}

Inspired by \textsc{ReAct}~\citep{yao2023react}, we introduce \textsc{ReAct-BM25}, which
runs \textsc{BM25} over several rounds instead of once. An LLM (in our
experiments, \texttt{gpt-oss-120b}) plays two roles: as \emph{planner} it writes
the search query, and as \emph{observer} it reads the retrieved articles (from the planner's query) and keeps
only those that help answer the question. The kept articles are passed back to the
planner, which writes a new query for the next round. We investigate if this
iterative process can narrow the performance gap between sparse and dense retrieval.
The full prompts are shown in Figures~\ref{fig:react_planner_system_prompt},
\ref{fig:react_planner_initial_prompt}, \ref{fig:react_planner_followup_prompt},
and~\ref{fig:react_observer_prompt}.

Concretely, in the first round, the planner receives the original \texttt{Question + Facts} query and produces a Greek legal keyword query for \textsc{BM25-GreekStemmer}, much as in LLM query reformulation (\S\ref{subsec:llm_query_reformulation}). BM25 retrieves the top 100 candidate articles, and the observer reviews them against the original query, retaining those that could provide useful evidence. In subsequent rounds, the planner inspects the original and previous queries, along with the articles retained by the observer. It then produces a new keyword query that explores a different legal angle or uses alternative terminology.

We run this process for up to 10 rounds. The observer returns binary decisions (regarding which retrieved articles to retain), so we need a 
way to convert the retained articles into a ranked list. We sort the retained articles by three keys: first, the number of rounds in which the observer retained the article; second, the earliest round in which it was retained; and third, its best BM25 rank 
from all the rounds where it was retained. The first key is the primary one. Articles with the same first-key values, are ranked by the second key; and articles with the same second-key value, are ranked by the third key. Thus, an article retained in multiple rounds (with multiple versions of the query) is ranked above an article retained only once. Among articles retained equally often, earlier retention is preferred; the intuition is that later rounds may have introduced query drift. BM25 rank is used only as the final tie-breaker. 
We retain up to 100 articles in the final output and report results after rounds 1, 2, 3, 5, and 10, with direct BM25 retrieval (without any reformulation rounds) reported as round 0.

\section{Experimental Results}
\label{sec:results}

Table~\ref{tab:main_results} reports 
our main results. Additional results are presented in Tables~\ref{tab:interventions} and \ref{tab:cost_quality}, discussed below. More detailed results are reported in Appendix~\ref{app:detailed_results}.

\subsection{Results of Vanilla Retrievers}
\label{subsec:results_main}

Table~\ref{tab:main_results} shows that most dense retrievers outperform BM25 on Recall@100, our main evaluation metric (\S\ref{subsec:evaluation_metrics}); we use \textsc{BM25-GreekStemmer} here, whose performance is indistinguishable from the second best BM25 variant (Table~\ref{tab:app_oneshot_sparse_dense}). \texttt{Gemini-001} achieves the highest Recall@100 (0.77), compared with 0.36 for BM25. Among the locally hosted dense models, \texttt{Euler-Legal-V1} reaches 0.68 and \texttt{Qwen3-8B}  0.67. Within the Qwen3 family, the larger variants achieve higher coverage: 0.67 for 8B, 0.62 for 4B, and 0.45 for 0.6B. \texttt{Nomic-v1.5} is a clear outlier, with 0.08 Recall@100, possibly because it was trained only on English data rather than on multilingual data.

\begin{table}[t]
\centering
\footnotesize
\setlength{\tabcolsep}{1.6pt}
\renewcommand{\arraystretch}{0.97}
\begin{tabular}{@{}>{\raggedright\arraybackslash}p{0.335\linewidth}ccccc@{}}
\toprule
\textbf{Model} & \textbf{N@10} & \textbf{N@100} & \textbf{R@10} & \textbf{R@100} & \textbf{MAP} \\
\midrule
\rowcolor{headerbg}
\multicolumn{6}{@{}l}{\textit{Dense retrieval, no query reformulation}} \\
\texttt{Gemini-001}      & 0.39 & \textbf{0.47} & 0.48 & \textbf{0.77} & 0.33 \\
\texttt{Qwen3-8B}        & 0.28 & 0.36 & 0.38 & 0.67 & 0.24 \\
\texttt{Euler-Legal-V1}        & 0.25 & 0.34 & 0.34 & 0.68 & 0.21 \\
\texttt{Qwen3-4B}        & 0.24 & 0.32 & 0.31 & 0.62 & 0.20 \\
\texttt{Jina-v5-small}       & 0.22 & 0.29 & 0.29 & 0.57 & 0.18 \\
\texttt{Arctic-v2} (0.6B)       & 0.22 & 0.28 & 0.28 & 0.51 & 0.18 \\
\texttt{Qwen3-0.6B}      & 0.14 & 0.20 & 0.20 & 0.45 & 0.11 \\
\texttt{EmbGemma-300M}        & 0.10 & 0.16 & 0.15 & 0.38 & 0.08 \\
\texttt{Nomic}-v1.5 (137M)      & 0.01 & 0.02 & 0.01 & 0.08 & 0.01 \\
\midrule
\rowcolor{headerbg}
\multicolumn{6}{@{}l}{\textit{Sparse retrieval, no query reformulation}} \\
\textsc{BM25}            & 0.10 & 0.14 & 0.16 & 0.36 & 0.09 \\
\midrule
\rowcolor{headerbg}
\multicolumn{6}{@{}l}{\textit{LLM query reformulation}} \\
Reform-\texttt{Qwen3-8B} & 0.33 & 0.41 & 0.44 & \localbest{0.73} & 0.28 \\
Reform-\textsc{BM25}     & 0.20 & 0.28 & 0.30 & 0.60 & 0.16 \\
\midrule
\rowcolor{headerbg}
\multicolumn{6}{@{}l}{\textit{Multi-round (iterative) retrieval}} \\
\textsc{ReAct-BM25}      & \best{0.43} & \best{0.47} & \best{0.52} & 0.67 & \best{0.37} \\
\bottomrule
\end{tabular}
\vspace{-2mm}
\caption{\textbf{Main results} on GreekBarRetrieval. 
N@$k$ is nDCG@$k$, R@$k$ is Recall@$k$, MAP is MAP@100. \textsc{BM25} denotes \textsc{BM25-GreekStemmer} with default parameters, the representative BM25 variant (see also Table~\ref{tab:app_oneshot_sparse_dense}).
Reform- denotes query reformulation. Reform- and \textsc{ReAct-} use \texttt{gpt-oss-120b} for query reformulation and as planner/observer, respectively. Bold marks the best overall score. Blue shading marks the best score among locally hosted systems. 
Most considered differences are \textbf{statistically significant} (Appendix~\ref{app:significance_tests}, Table~\ref{tab:significance_tests}), with  exceptions in \textsc{ReAct-BM25} vs.\ Reform-\texttt{Qwen3-8B}, and \textsc{ReAct-BM25} vs.\ \texttt{Gemini-001} . 
}
\vspace{-5mm}
\label{tab:main_results}
\end{table}

\subsection{Query Reformulation Results}
\label{subsec:results_query_formulation}

As shown in Table~\ref{tab:main_results}, LLM reformulation benefits sparse retrieval more than dense retrieval: it raises \textsc{BM25} Recall@100 from 0.36 to 0.60, compared with an increase from 0.67 to 0.73 for \texttt{Qwen3-8B}. These improvements are statistically significant on all five metrics after correction for multiple comparisons (Table~\ref{tab:significance_tests}, Appendix~\ref{app:significance_tests}). The detailed results in Table~\ref{tab:app_interventions} (Appendix~\ref{app:results_interventions}) show the same pattern for the other evaluated retrievers: all three sparse variants gain at least 0.20 with query reformulation, whereas the seven dense encoders, excluding  \texttt{Nomic-v1.5}, gain between 0.03 and 0.09. Reform-\texttt{Qwen3-8B} achieves the highest Recall@100 among the locally hosted systems.

\subsection{ReAct-BM25 Results}
\label{subsec:results_agentic_bm25}

\textsc{ReAct-BM25} further improves the performance of BM25, reaching 0.67 Recall@100 (Table~\ref{tab:main_results}), compared with 0.60 for Reform-\textsc{BM25}. We conjecture that this additional gain comes from the feedback loop: unlike one-shot reformulation, \textsc{ReAct-BM25} reviews the retrieved articles and subsequently uses them to guide the next query. Interestingly, \textsc{ReAct-BM25} reaches the same Recall@100 as vanilla \texttt{Qwen3-8B} (0.67); repeated LLM-guided query reformulation may help BM25 capture some of the semantic, non-surface matches as in dense retrieval. \textsc{ReAct-BM25} also obtains the highest nDCG@10 (0.43), nDCG@100 (0.47), Recall@10 (0.52), and MAP@100 (0.37) 
scores among all tested retrievers. These results show that \textsc{ReAct-BM25} effectively ranks relevant articles near the top and may therefore benefit RAG systems with smaller context windows.

Although \textsc{ReAct-BM25} does not surpass Reform-\texttt{Qwen3-8B} in Recall@100 (0.67 vs.\ 0.73), we did not find the difference to be statistically significant (Table~\ref{tab:significance_tests}, Appendix~\ref{app:significance_tests}). On the other hand, \textsc{ReAct-BM25} outperforms Reform-\texttt{Qwen3-8B} in nDCG@10, nDCG@100, MAP@100, and these differences \emph{are} statistically significant (Table~\ref{tab:significance_tests}); we did not test the significance of the difference in Recall@10, where the gap is also large, in favor of \textsc{ReAct-BM25}. 

The difference in Recall@100 between \texttt{Gemini-001} (0.77) and \textsc{ReAct-BM25} (0.67) is statistically significant (Table~\ref{tab:significance_tests}), but we did not detect a statistically significant difference in their nDCG@10, nDCG@100, MAP@100 scores. 


The performance of \textsc{ReAct-BM25} improves at every reported round  (Table~\ref{tab:app_agentic_bm25_metrics}, App.~\ref{app:results_agentic_bm25}).
Recall@100 increases from 0.45 at round 1 to 0.52 at round 2, 0.55 at round 3, 0.62 at round 5, and 0.67 at round 10. nDCG@10, nDCG@100, Recall@10, and MAP@100 follow the same pattern. 
Most of the improvement in these  metrics occurs in the first round, while later rounds continue to improve Recall@100. Further analysis (Table~\ref{tab:react_candidate_recall}, App.~\ref{app:results_agentic_bm25}) shows the observer excludes some retrieved relevant articles, limiting the final Recall@100 of \textsc{ReAct-BM25} to 0.67 (Table~\ref{tab:main_results}). 

\subsection{Tuning, PRF, RRF, Translation Results}
\label{subsec:results_comparison}

Table~\ref{tab:interventions} shows that BM25 parameter tuning (\S\ref{sec:bm25_parameter_tuning}), pseudo-relevance feedback (PRF, \S\ref{subsec:pseudo_relevance_feedback}), fusion (RRF, \S\ref{subsec:sparse_dense_fusion}), and English translation  (\S\ref{subsec:english_translation_baseline})  all have negligible effects on \textsc{BM25-GreekStemmer}, one of the two best and effectively indistinguishable BM25 variants, and on the best locally hosted dense retriever (\texttt{Qwen3-8B}). More detailed results, with similar findings, are reported in Appendices~\ref{app:bm25_tuning} and~\ref{app:results_interventions}. The gain from query reformulation is substantially larger  (Table~\ref{tab:interventions}). This advantage is most pronounced for BM25, where reformulation improves Recall@100 by $+0.23$, compared with $+0.02$ for translation; for dense retrieval, the gain from reformulation ($+0.06$) is comparable to that from translation ($+0.04$), indicating that reformulation is useful but less impactful than in BM25.

\begin{table}[hbt]
\centering
\footnotesize
\setlength{\tabcolsep}{3pt}
\renewcommand{\arraystretch}{0.97}
\begin{tabular}{@{}>{\raggedright\arraybackslash}p{0.60\linewidth}cc@{}}
\toprule
\textbf{Retriever} & \textbf{R@100} & \textbf{$\Delta$} \\
\midrule
\rowcolor{headerbg}
\textit{Sparse:} \textsc{BM25-GreekStemmer} & 0.36 & ---\\
\quad + parameter tuning & 0.41 & $+0.05$ \\
\quad + pseudo-relevance feedback & 0.37 & $+0.01$ \\
\quad + English translation\textsuperscript{$\dagger$} & 0.38 & $+0.02$ \\
\quad + LLM reformulation & 0.60 & $+0.23$ \\
\quad + LLM reformulation + tuning & 0.61 & $+0.25$ \\
\midrule
\rowcolor{headerbg}
\textit{Dense:} \texttt{Qwen3-8B} & 0.67 & --- \\
\quad + pseudo-relevance feedback & 0.66 & $-0.01$ \\
\quad + English translation & 0.71 & $+0.04$ \\
\quad + fusion with \textsc{BM25} (RRF) & 0.64 & $-0.03$ \\
\quad + LLM reformulation & 0.73 & $+0.06$ \\
\bottomrule
\end{tabular}
\vspace{-2mm}
\caption{The effect of \textbf{BM25 parameter tuning}, \textbf{pseudo-relevance feedback} (PRF), \textbf{fusion} (RRF), English \textbf{translation}, and \textbf{LLM query reformulation} on the best BM25 variant (\textsc{BM25-GreekStemmer}) and the best locally hosted dense retriever. 
\textsuperscript{$\dagger$}\textsc{BM25-spaCy} used, to have  comparable (\textsc{spaCy}) pre-processing in both languages. 
Comparing \textsc{BM25-GreekStemmer} against the English \textsc{BM25-spaCy} instead gives $-0.11$. 
}
\vspace{-5mm}
\label{tab:interventions}
\end{table}

\subsection{Retrieval Performance vs.\ Inference Cost}
\label{subsec:results_cost}

Table~\ref{tab:cost_quality} explores the tradeoff between retrieval performance and inference cost. It reports nDCG@10 and Recall@100 as measures of top-rank quality and overall evidence retrieval, respectively. It also reports per-query LLM calls, tokens, and wall-clock time in seconds. The systems lead on different metrics. Reform-\texttt{Qwen3-8B} reaches 0.73 Recall@100 and 0.33 nDCG@10 with two calls and 2.6k rewrite-generate tokens plus 0.3k rewritten-query embedding tokens per query. Ten-round \textsc{ReAct-BM25} reaches 0.67 Recall@100 and 0.43 nDCG@10 with 20 calls and 617.9k tokens per query, i.e., with substantially larger inference cost;  the observer accounts for $84\%$ of these tokens. All calls in Table~\ref{tab:cost_quality} were served locally, so these are  compute costs rather than paid API costs.\footnote{Model inference was served on a Mac Studio with an Apple M3 Ultra, 512~GB of unified memory.}


\begin{table}[ht]
\centering
\footnotesize
\setlength{\tabcolsep}{2.2pt}
\renewcommand{\arraystretch}{0.97}

\begin{tabular}{@{}>{\raggedright\arraybackslash}p{0.30\linewidth}ccccc@{}}
\toprule
\textbf{System} &
\textbf{N@10} &
\textbf{R@100} &
\textbf{Calls/q} &
\textbf{Tok/q} &
\textbf{Time/q} \\
\midrule

\textsc{BM25}
& 0.10 & 0.36 & 0 & --- & 0.43 \\

\texttt{Qwen3-8B}
& 0.28 & 0.67 & 1 & 1.5k & 1.23 \\

Reform-\texttt{Qwen3-8B}
& 0.33 & \best{0.73} & 2 & 2.9k & 5.13 \\

\midrule
\rowcolor{headerbg}
\multicolumn{6}{@{}l}{\textsc{ReAct-BM25}} \\

\quad 1 round
& 0.34 & 0.45 & 2 & 54.9k & 141.9 \\

\quad 2 rounds
& 0.38 & 0.52 & 4 & 111.9k & 289.3 \\

\quad 3 rounds
& 0.40 & 0.55 & 6 & 170.9k & 442.0 \\

\quad 5 rounds
& 0.41 & 0.62 & 10 & 294.2k & 760.8 \\

\quad 10 rounds
& \best{0.43} & 0.67 & 20 & 617.9k & 1597.7 \\

\bottomrule
\end{tabular}
\vspace{-2mm}
\caption{\textbf{Retrieval performance} (nDCG@10, Recall@100) and \textbf{inference cost} (LLM calls/query, tokens/query, time/query in sec.).
Tok/q reports query tokens including both rewrite or planner-observer LLM tokens and embedding tokens. Times for 
intermediate rounds of \textsc{ReAct-BM25} are token-proportional
estimates from the ten-round time.}
\vspace{-5mm}
\label{tab:cost_quality}
\end{table}

The systems also differ in their infrastructure requirements. Because a \textsc{BM25} index is built from raw tokens, it needs no GPU to encode the retrieval pool, no vector database, and no re-indexing when an embedding model is replaced. Articles can be added or amended by updating the index. \textsc{ReAct-BM25} keeps this property at the expense of increased inference time. By contrast, the dense systems reverse the arrangement, paying a lot to build the index and less per query at inference time.

\section{Related Work}
\label{sec:related}

\paragraph{Legal retrieval and RAG benchmarks.}
Recent legal NLP work increasingly evaluates retrieval as a key component of legal RAG. LegalBench-RAG~\citep{pipitone2024legalbenchrag} focuses on retrieving legally relevant snippets for grounded legal answering, while \citet{zheng2025reasoningfocused} introduce retrieval tasks designed around legal reasoning needs, including Bar Exam QA and Housing Statute QA. The Massive Legal Embedding Benchmark (MLEB) comprises ten expert-annotated evaluation sets, including a U.S. Bar Exam QA task for retrieving relevant cases and legal literature~\citep{butler2025mleb}.
Other resources study retrieval across jurisdictions and document types, including statutory retrieval in Belgian, Italian, and German law~\citep{louis2021bsard,noce-etal-2026-jurifindit,weber2025gerlerb}, COLIEE shared tasks~\citep{goebel2026coliee}, Thai legal QA~\citep{akarajaradwong-etal-2025-nitibench}, and U.S. precedent retrieval~\citep{mahari2024lepard}. Unlike these benchmarks, GreekBarRetrieval focuses on article-level statutory retrieval for Greek bar-exam questions, using the statutory articles cited in official solutions as gold labels.

\paragraph{GreekBarBench and Greek  NLP.}
GreekBarRetrieval is derived from GreekBarBench~\citep{chlapanis-etal-2025-greekbarbench}, which evaluates LLMs on Greek bar exam questions requiring free-text legal reasoning and citations. However, GreekBarBench does not involve true retrieval; instead, each question is accompanied by the gold relevant articles and distractors. We instead focus on the retrieval step, whether a system can recover the statutory articles needed before answer generation. This separates retrieval from reasoning or generation failures.

Greek remains less resourced than English in NLP~\citep{papantoniou2024nlp}, and its morphology makes lexical matching harder, motivating Greek-specific normalization and stemming, including the stemmer of \citet{ntais2006development}.

\paragraph{Sparse, dense, hybrid retrieval.}
Sparse and dense retrieval capture different relevance signals. BM25 remains the standard sparse baseline~\citep{robertson2009bm25} and is especially relevant in law, where exact statutory terms, article references, doctrinal expressions etc.\ may need to be matched exactly. Dense retrieval instead helps when the query and relevant authorities express the same legal concepts with different wordings. Hence, legal retrieval systems are often hybrid, i.e., 
they combine sparse and dense retrieval~\citep{shao2020thuir,nigam2022coliee}. We evaluated Greek-aware BM25 variants, dense embedding models, and Reciprocal Rank Fusion (RRF)~\citep{cormack2009rrf}, a rank-based method for combining retrievers without score calibration.

\paragraph{Query reformulation.}
Legal query reformulation aims to reduce the mismatch between user queries and legal authorities. GuRE~\citep{kim-etal-2025-gure} applies generative query rewriting to legal passage retrieval. \citet{zhou2023queryselection} study a related problem in legal case retrieval, showing that long legal queries often contain noisy details and benefit from selecting legally salient content. Rewrite-Retrieve-Read~\citep{ma2023rewriteretrieveread} uses an LLM to reformulate queries before retrieval in RAG pipelines. We followed this direction, but focused on Greek bar-exam questions, facts, and statutory article retrieval rather than case or passage retrieval.

\paragraph{Translation and multilingual retrieval.}
Translation is often used to adapt retrieval methods to lower-resource languages. Prior work has compared query and document translation for cross-lingual retrieval~\citep{saleh2020documenttranslation}, while MIRACL~\citep{zhang2023miracl} highlights the need for multilingual retrieval evaluation. Our setting is not cross-lingual; both queries and articles are originally Greek, and translation is used only as a baseline alternative to Greek retrieval.

\paragraph{Iterative and agentic retrieval.}
ReAct~\citep{yao2023react} and IRCoT~\citep{trivedi-etal-2023-interleaving} show retrieval can be interleaved with reasoning instead of performed as a single step. This matters in legal retrieval, where a question may require several articles to be searched from different legal angles.
We introduced and evaluated \textsc{ReAct-BM25}, where retrieved articles guide later query reformulations across multiple search rounds (\S\ref{subsec:agentic_bm25_retrieval}).

\paragraph{Pseudo-relevance feedback.}
Relevance feedback updates a query using user feedback for retrieved documents. Rocchio's method 
is the classical reference~\citep{rocchio1971relevance}. Pseudo-relevance feedback (PRF) removes the need for user feedback by treating the top-ranked documents from an initial retrieval pass as relevant. This idea has also been adapted to dense retrieval, for example in ColBERT-PRF~\citep{wang2021colbertprf}. PRF is relevant to our work because, like LLM-based reformulation, it modifies the query before a second retrieval pass, but it relies only on initially retrieved documents rather than an explicit query reformulation.

\section{Discussion}
\label{sec:discussion}

Vanilla dense retrieval performs better than sparse retrieval (Table~\ref{tab:main_results}), but this result does not make sparse retrieval generally unsuitable for legal retrieval. The two approaches rely on different signals. Dense retrievers can match queries with statutory articles even when they use different wordings. Sparse retrieval is more effective when terms need to be matched exactly (e.g., article references, statute names, particular doctrinal expressions). 
Simply merging the rankings of a sparse and a dense retriever, however, as in RRF (\S\ref{subsec:sparse_dense_fusion}), provides no benefit in GreekBarRetrieval (Table~\ref{tab:interventions}).

By contrast, LLM query reformulation improves both sparse and dense retrievers (Table~\ref{tab:main_results}).
The improvement is particularly strong for BM25, presumably because reformulation facilitates matches between semantically equivalent query and statutory terms, bringing to BM25 some of the benefits of dense retrieval. Reformulation also benefits both sparse and dense retrievers by removing from the query irrelevant case facts. These findings indicate that retrieval performance depends not only on the retrieval model, but also on how clearly the query represents the information need in legal  terms.

\textsc{ReAct-BM25} produces the strongest results for nDCG@10, nDCG@100, Recall@10, MAP@100, placing relevant articles more consistently near the top of the ranking, and allowing conventional, less computationally intensive (compared to embedding models) inverted term indices to be employed. However, \textsc{ReAct-BM25} does not exceed LLM reformulated dense retrieval in Recall@100. 
It is, therefore, more useful for obtaining a small set of relevant articles than for maximizing overall evidence coverage. More importantly, the improvement in top-position ranking comes at a substantial computational cost (\S\ref{subsec:results_cost}). Table~\ref{tab:cost_analysis} (App.~\ref{app:cost_analysis}) shows that 
Reform-\texttt{Qwen3-8B} uses one reformulation and one embedding call, processing approx.\ 2.6k LLM tokens and 0.3k rewritten-query embedding token per query. The ten-round \textsc{ReAct-BM25} system uses 20 planner and observer calls and processes approx.\ 617.9k LLM tokens per query. It therefore processes about 240 times more LLM tokens and has an estimated cost per query about 154 times higher. Its sequential calls also result in much greater latency. By comparison, embedding the full retrieval pool with 
\texttt{Qwen3-8B} 
costs only \$0.0385.

These findings favor query reformulation followed by dense retrieval (e.g., Reform-\texttt{Qwen3-8B}, Table~\ref{tab:main_results}) as the practical default when both evidence coverage and computational cost matter. Iterative sparse retrieval (as in \textsc{ReAct-BM25}) may be useful when only a few articles can be passed to a downstream model, when a conventional sparse retriever has to be used, or when a difficult query (e.g., with insufficient evidence from reformulated dense retrieval) justifies additional computation. 

\section{Conclusions}
\label{sec:conclusions}

We introduced \textbf{GreekBarRetrieval}, a public benchmark linking 283 Greek bar exam questions to the statutory articles cited in their official solutions within a retrieval pool of 6,308 candidates. The new benchmark complements GreekBarBench, which did not include retrieval.
Experimenting with three BM25 variants and nine dense retrievers, we found vanilla dense retrieval to far outperform vanilla sparse retrieval in overall evidence coverage. However, LLM query reformulation helps \textsc{BM25} close the gap, by bringing to sparse retrieval some of the inexact mapping benefits of dense retrieval. Reformulation also helps sparse and dense retrievers discard irrelevant case facts. A \textsc{ReAct}-like multi-round reformulation that we introduced helps BM25 obtain the best performance at top ranking positions, at the expense of substantially increased inference cost. 
Compared with BM25 parameter tuning, PRF, RRF, and English translation, query reformulation is by far the most effective enhancement of sparse retrieval, while dense retrieval also benefits, obtaining the strongest overall coverage. 

\section*{Limitations}
\label{sec:limitations}

GreekBarRetrieval evaluates systems against the statutory articles cited in the official Greek bar bench examination solutions. 
This ground truth is not exhaustive; further relevant articles may exist and are not credited by our evaluation.

We report paired significance tests for selected comparisons. With 283 queries and a mean of 2.74 relevant articles per query, small differences should be treated cautiously, even when statistically significant. All LLM-dependent results come from a single generation path per query. Translation, both reformulation prompts, and \textsc{ReAct-BM25} were each run once, due to limited computational resources, so we cannot report run-to-run variance.

The BM25 parameter sweep was optimized on the full benchmark and is reported only as an upper bound. GreekBarRetrieval does not currently provide a development subset, which future work should add to facilitate hyper-parameter tuning.

Our claims that reformulation brings some of the inexact matching benefits of dense retrieval to sparse retrieval,  while also helping both types of retrievers remove irrelevant case facts,  rest on indirect evidence, mostly the much larger gains for sparse retrieval, but also the fact that both dense and sparse retrievers improve. Appendix~\ref{app:qualitative_reformulation_example} provides an indicative example, but a broader analysis is needed to solidify these claims.

We did not include any re-rankers in our baselines. Since the main advantages of \textsc{ReAct-BM25} are improved performance at top ranking positions and its ability to use conventional sparse retrievers, comparing it against baselines, especially sparse ones, coupled with re-rankers would be particularly interesting and might diminish those advantages. 

The benchmark focuses on Greek statutory article retrieval for bar-exam-style questions. It does not cover case law, secondary sources, dynamic legal corpora, or real user search sessions.

Finally, we evaluate retrieval independently of its effect on end-to-end legal QA. Future work should test whether the observed retrieval gains lead to more accurate and better grounded answers.

\bibliography{custom}

@article{martinezgil2023survey,
    title = {A survey on legal question–answering systems},
    journal = {Computer Science Review},
    volume = {48},
    pages = {100552},
    year = {2023},
    issn = {1574-0137},
    doi = {https://doi.org/10.1016/j.cosrev.2023.100552},
    url = {https://www.sciencedirect.com/science/article/pii/S1574013723000199},
    author = {Jorge Martinez-Gil}
}

@book{manning2008introduction,
  author={Manning, Christopher D. and Raghavan, Prabhakar and Schütze, Hinrich},
  title     = {Introduction to Information Retrieval},
  year      = {2008},
  publisher = {Cambridge University Press},
  address   = {Cambridge, UK},
  isbn      = {9780521865715},
  doi       = {10.1017/CBO9780511809071},
  url       = {https://www.cambridge.org/highereducation/books/introduction-to-information-retrieval/669D108D20F556C5C30957D63B5AB65C}
}

@inproceedings{su-etal-2024-stard,
    title = "{STARD}: A {C}hinese Statute Retrieval Dataset Derived from Real-life Queries by Non-professionals",
    author = "Su, Weihang  and
      Hu, Yiran  and
      Xie, Anzhe  and
      Ai, Qingyao  and
      Bing, Quezi  and
      Zheng, Ning  and
      Liu, Yun  and
      Shen, Weixing  and
      Liu, Yiqun",
    editor = "Al-Onaizan, Yaser  and
      Bansal, Mohit  and
      Chen, Yun-Nung",
    booktitle = "Findings of the Association for Computational Linguistics: EMNLP 2024",
    month = nov,
    year = "2024",
    address = "Miami, Florida, USA",
    publisher = "Association for Computational Linguistics",
    url = "https://aclanthology.org/2024.findings-emnlp.625/",
    doi = "10.18653/v1/2024.findings-emnlp.625",
    pages = "10658--10671"
}

@inproceedings{ma2021lecard,
author = {Ma, Yixiao and Shao, Yunqiu and Wu, Yueyue and Liu, Yiqun and Zhang, Ruizhe and Zhang, Min and Ma, Shaoping},
title = {LeCaRD: A Legal Case Retrieval Dataset for Chinese Law System},
year = {2021},
isbn = {9781450380379},
publisher = {Association for Computing Machinery},
address = {New York, NY, USA},
url = {https://doi.org/10.1145/3404835.3463250},
doi = {10.1145/3404835.3463250},
booktitle = {Proceedings of the 44th International ACM SIGIR Conference on Research and Development in Information Retrieval},
pages = {2342–2348},
numpages = {7},
location = {Virtual Event, Canada},
series = {SIGIR '21}
}

@inproceedings{li2024lecardv2,
author = {Li, Haitao and Shao, Yunqiu and Wu, Yueyue and Ai, Qingyao and Ma, Yixiao and Liu, Yiqun},
title = {LeCaRDv2: A Large-Scale Chinese Legal Case Retrieval Dataset},
year = {2024},
isbn = {9798400704314},
publisher = {Association for Computing Machinery},
address = {New York, NY, USA},
url = {https://doi.org/10.1145/3626772.3657887},
doi = {10.1145/3626772.3657887},
booktitle = {Proceedings of the 47th International ACM SIGIR Conference on Research and Development in Information Retrieval},
pages = {2251–2260},
numpages = {10},
location = {Washington DC, USA},
series = {SIGIR '24}
}

@inproceedings{joshi-etal-2024-il,
    title = "{IL}-{TUR}: Benchmark for {I}ndian Legal Text Understanding and Reasoning",
    author = "Joshi, Abhinav  and
      Paul, Shounak  and
      Sharma, Akshat  and
      Goyal, Pawan  and
      Ghosh, Saptarshi  and
      Modi, Ashutosh",
    editor = "Ku, Lun-Wei  and
      Martins, Andre  and
      Srikumar, Vivek",
    booktitle = "Proceedings of the 62nd Annual Meeting of the Association for Computational Linguistics (Volume 1: Long Papers)",
    month = aug,
    year = "2024",
    address = "Bangkok, Thailand",
    publisher = "Association for Computational Linguistics",
    url = "https://aclanthology.org/2024.acl-long.618/",
    doi = "10.18653/v1/2024.acl-long.618",
    pages = "11460--11499"
}

@article{butler2025mleb,
  title   = {The Massive Legal Embedding Benchmark (MLEB)},
  author  = {Butler, Umar and Butler, Abdur-Rahman and Malec, Adrian Lucas},
  journal = {arXiv preprint arXiv:2510.19365},
  year    = {2025}
}

@misc{papantoniou2024nlp,
  title         = {NLP for The Greek Language: A Longer Survey},
  author        = {Papantoniou, Katerina and Tzitzikas, Yannis},
  year          = {2024},
  eprint        = {2408.10962},
  archivePrefix = {arXiv},
  primaryClass  = {cs.CL},
  url           = {https://arxiv.org/abs/2408.10962}
}

@inproceedings{chlapanis-etal-2025-greekbarbench,
    title = "{G}reek{B}ar{B}ench: A Challenging Benchmark for Free-Text Legal Reasoning and Citations",
    author = "Chlapanis, Odysseas S.  and
      Galanis, Dimitrios  and
      Aletras, Nikolaos  and
      Androutsopoulos, Ion",
    editor = "Christodoulopoulos, Christos  and
      Chakraborty, Tanmoy  and
      Rose, Carolyn  and
      Peng, Violet",
    booktitle = "Findings of the Association for Computational Linguistics: EMNLP 2025",
    month = nov,
    year = "2025",
    address = "Suzhou, China",
    publisher = "Association for Computational Linguistics",
    url = "https://aclanthology.org/2025.findings-emnlp.1368/",
    doi = "10.18653/v1/2025.findings-emnlp.1368",
    pages = "25099--25119",
    ISBN = "979-8-89176-335-7"
}

@inproceedings{kim-etal-2025-gure,
    title = "{G}u{RE}:Generative Query {RE}writer for Legal Passage Retrieval",
    author = "Kim, Daehui  and
      Kang, Deokhyung  and
      Kim, Jonghwi  and
      Ryu, Sangwon  and
      Lee, Gary",
    editor = "Aletras, Nikolaos  and
      Chalkidis, Ilias  and
      Barrett, Leslie  and
      Goanț{\u{a}}, C{\u{a}}t{\u{a}}lina  and
      Preoțiuc-Pietro, Daniel  and
      Spanakis, Gerasimos",
    booktitle = "Proceedings of the Natural Legal Language Processing Workshop 2025",
    month = nov,
    year = "2025",
    address = "Suzhou, China",
    publisher = "Association for Computational Linguistics",
    url = "https://aclanthology.org/2025.nllp-1.31/",
    doi = "10.18653/v1/2025.nllp-1.31",
    pages = "424--438",
    ISBN = "979-8-89176-338-8"
}

@inproceedings{noce-etal-2026-jurifindit,
    title = "{J}uri{F}ind{IT}: an {I}talian legal retrieval dataset",
    author = "Noce, Niko Dalla  and
      Colla, Davide  and
      Doust, Sina Farhang  and
      De Mattei, Lorenzo  and
      Bacciu, Davide",
    editor = "Demberg, Vera  and
      Inui, Kentaro  and
      Marquez, Llu{\'i}s",
    booktitle = "Findings of the {A}ssociation for {C}omputational {L}inguistics: {EACL} 2026",
    month = mar,
    year = "2026",
    address = "Rabat, Morocco",
    publisher = "Association for Computational Linguistics",
    url = "https://aclanthology.org/2026.findings-eacl.221/",
    doi = "10.18653/v1/2026.findings-eacl.221",
    pages = "4223--4241",
    ISBN = "979-8-89176-386-9"
}

@mastersthesis{ntais2006development,
  author  = {Ntais, Georgios},
  title   = {Development of a Stemmer for the {Greek} Language},
  school  = {Stockholm University / Royal Institute of Technology},
  address = {Stockholm, Sweden},
  month   = feb,
  year    = {2006},
  url     = {https://people.dsv.su.se/~hercules/papers/Ntais_greek_stemmer_thesis_final.pdf}
}

@misc{pipitone2024legalbenchrag,
      title={LegalBench-RAG: A Benchmark for Retrieval-Augmented Generation in the Legal Domain}, 
      author={Nicholas Pipitone and Ghita Houir Alami},
      year={2024},
      eprint={2408.10343},
      archivePrefix={arXiv},
      primaryClass={cs.AI},
      url={https://arxiv.org/abs/2408.10343}, 
}

@inproceedings{zheng2025reasoningfocused,
author = {Zheng, Lucia and Guha, Neel and Arifov, Javokhir and Zhang, Sarah and Skreta, Michal and Manning, Christopher D. and Henderson, Peter and Ho, Daniel E.},
title = {A Reasoning-Focused Legal Retrieval Benchmark},
year = {2025},
isbn = {9798400714214},
publisher = {Association for Computing Machinery},
address = {New York, NY, USA},
url = {https://doi.org/10.1145/3709025.3712219},
doi = {10.1145/3709025.3712219},
booktitle = {Proceedings of the 2025 Symposium on Computer Science and Law},
pages = {169–193},
numpages = {25},
location = {Munich, Germany},
series = {CSLAW '25}
}

@inproceedings{akarajaradwong-etal-2025-nitibench,
    title = "{N}iti{B}ench: Benchmarking {LLM} Frameworks on {T}hai Legal Question Answering Capabilities",
    author = "Akarajaradwong, Pawitsapak  and
      Pothavorn, Pirat  and
      Chaksangchaichot, Chompakorn  and
      Tasawong, Panuthep  and
      Nopparatbundit, Thitiwat  and
      Pratai, Keerakiat  and
      Nutanong, Sarana",
    editor = "Christodoulopoulos, Christos  and
      Chakraborty, Tanmoy  and
      Rose, Carolyn  and
      Peng, Violet",
    booktitle = "Proceedings of the 2025 Conference on Empirical Methods in Natural Language Processing",
    month = nov,
    year = "2025",
    address = "Suzhou, China",
    publisher = "Association for Computational Linguistics",
    url = "https://aclanthology.org/2025.emnlp-main.1739/",
    doi = "10.18653/v1/2025.emnlp-main.1739",
    pages = "34304--34327",
    ISBN = "979-8-89176-332-6"
}

@article{goebel2026coliee,
  title   = {The {COLIEE} 2025 Competition on Legal Information Extraction and Entailment: Overview, Discussion, and Dataset Expansion},
  author  = {Goebel, Randy and Kano, Yoshinobu and Kim, Mi-Young and Kwan, Calum and Rabelo, Juliano and Satoh, Ken and Yamada, Hiroaki and Yoshioka, Masaharu},
  journal = {The Review of Socionetwork Strategies},
  volume  = {20},
  number  = {1},
  pages   = {183--213},
  year    = {2026},
  doi     = {10.1007/s12626-026-00199-9},
  url     = {https://doi.org/10.1007/s12626-026-00199-9}
}

@inproceedings{
    yao2023react,
    title={ReAct: Synergizing Reasoning and Acting in Language Models},
    author={Shunyu Yao and Jeffrey Zhao and Dian Yu and Nan Du and Izhak Shafran and Karthik R Narasimhan and Yuan Cao},
    booktitle={The Eleventh International Conference on Learning Representations },
    year={2023},
    url={https://openreview.net/forum?id=WE_vluYUL-X}
}

@inproceedings{trivedi-etal-2023-interleaving,
    title = "Interleaving Retrieval with Chain-of-Thought Reasoning for Knowledge-Intensive Multi-Step Questions",
    author = "Trivedi, Harsh  and
      Balasubramanian, Niranjan  and
      Khot, Tushar  and
      Sabharwal, Ashish",
    editor = "Rogers, Anna  and
      Boyd-Graber, Jordan  and
      Okazaki, Naoaki",
    booktitle = "Proceedings of the 61st Annual Meeting of the Association for Computational Linguistics (Volume 1: Long Papers)",
    month = jul,
    year = "2023",
    address = "Toronto, Canada",
    publisher = "Association for Computational Linguistics",
    url = "https://aclanthology.org/2023.acl-long.557/",
    doi = "10.18653/v1/2023.acl-long.557",
    pages = "10014--10037"
}

@inproceedings{xu1996query,
  author    = {Xu, Jinxi and Croft, W. Bruce},
  title     = {Query Expansion Using Local and Global Document Analysis},
  booktitle = {Proceedings of the 19th Annual International {ACM} {SIGIR} Conference on Research and Development in Information Retrieval},
  series    = {{SIGIR} '96},
  pages     = {4--11},
  year      = {1996},
  publisher = {Association for Computing Machinery},
  doi       = {10.1145/243199.243202},
  url       = {https://doi.org/10.1145/243199.243202}
}

@inproceedings{louis2021bsard,
    title = "A Statutory Article Retrieval Dataset in {F}rench",
    author = "Louis, Antoine  and
      Spanakis, Gerasimos",
    editor = "Muresan, Smaranda  and
      Nakov, Preslav  and
      Villavicencio, Aline",
    booktitle = "Proceedings of the 60th Annual Meeting of the Association for Computational Linguistics (Volume 1: Long Papers)",
    month = may,
    year = "2022",
    address = "Dublin, Ireland",
    publisher = "Association for Computational Linguistics",
    url = "https://aclanthology.org/2022.acl-long.468/",
    doi = "10.18653/v1/2022.acl-long.468",
    pages = "6789--6803"
}

@incollection{weber2025gerlerb,
    author = "Weber, Malte and Paritala, Balaramakrishna and Rechu, Abhilash Reddy and Feddoul, Leila and Bonagiri, Suresh Kumar and Klewer, Norman and Karg, Pirmin Mathias and Unger, Christoph and Mauch, Marianne and König-Ries, Birgitta",
    title= {{G}er{L}e{RB} -- {G}erman Legislative Retrieval Benchmark},
    year = 2025,
    editor = "Gehlsen, Björn and Schnackenburg, André",
    doi = "10.18420/rvi2025-112",
    booktitle = "8. Fachtagung Rechts- und Verwaltungsinformatik (RVI 2025)",
    publisher = "Gesellschaft für Informatik e.V.",
    address = "Bonn",
    pissn = "2944-7682",
    pages = "157--168",
}

@inproceedings{mahari2024lepard,
    title = "{L}e{P}a{RD}: A Large-Scale Dataset of Judicial Citations to Precedent",
    author = "Mahari, Robert  and
      Stammbach, Dominik  and
      Ash, Elliott  and
      Pentland, Alex",
    editor = "Ku, Lun-Wei  and
      Martins, Andre  and
      Srikumar, Vivek",
    booktitle = "Proceedings of the 62nd Annual Meeting of the Association for Computational Linguistics (Volume 1: Long Papers)",
    month = aug,
    year = "2024",
    address = "Bangkok, Thailand",
    publisher = "Association for Computational Linguistics",
    url = "https://aclanthology.org/2024.acl-long.532/",
    doi = "10.18653/v1/2024.acl-long.532",
    pages = "9863--9877"
}

@article{robertson2009bm25,
    author = {Robertson, Stephen and Zaragoza, Hugo},
    title = {The Probabilistic Relevance Framework: BM25 and Beyond},
    journal = {Foundations and Trends in Information Retrieval},
    volume = {4},
    number = {1-2},
    pages = {1-174},
    year = {2009},
    month = {09},
    issn = {1554-0669},
    doi = {10.1561/1500000019},
    url = {https://doi.org/10.1561/1500000019},
    eprint = {https://www.emerald.com/ftinr/article-pdf/4/1-2/1/11486410/1500000019en.pdf},
}

@misc{shao2020thuir,
  title         = {{THUIR}@{COLIEE}-2020: Leveraging Semantic Understanding and Exact Matching for Legal Case Retrieval and Entailment},
  author        = {Shao, Yunqiu and Liu, Bulou and Mao, Jiaxin and Liu, Yiqun and Zhang, Min and Ma, Shaoping},
  year          = {2020},
  eprint        = {2012.13102},
  archivePrefix = {arXiv},
  primaryClass  = {cs.IR},
  doi           = {10.48550/arXiv.2012.13102},
  url           = {https://arxiv.org/abs/2012.13102},
  note          = {Presented at the Fourteenth International Workshop on Juris-Informatics ({JURISIN} 2020), {COLIEE} session}
}

@inproceedings{nigam2022coliee,
    author="Nigam, Shubham Kumar
    and Goel, Navansh
    and Bhattacharya, Arnab",
    editor="Takama, Yasufumi
    and Yada, Katsutoshi
    and Satoh, Ken
    and Arai, Sachiyo",
    title="nigam@COLIEE-22: Legal Case Retrieval and Entailment Using Cascading of Lexical and Semantic-Based Models",
    booktitle="New Frontiers in Artificial Intelligence",
    year="2023",
    publisher="Springer Nature Switzerland",
    address="Cham",
    pages="96--108",
    isbn="978-3-031-29168-5"
}

@inproceedings{cormack2009rrf,
  title     = {Reciprocal Rank Fusion Outperforms Condorcet and Individual Rank Learning Methods},
  author    = {Cormack, Gordon V. and Clarke, Charles L. A. and Buettcher, Stefan},
  booktitle = {Proceedings of the 32nd International ACM SIGIR Conference on Research and Development in Information Retrieval},
  series    = {SIGIR '09},
  pages     = {758--759},
  year      = {2009},
  publisher = {Association for Computing Machinery},
  address   = {New York, NY, USA},
  doi       = {10.1145/1571941.1572114},
  url       = {https://doi.org/10.1145/1571941.1572114}
}

@inproceedings{zhou2023queryselection,
    author = {Zhou, Youchao and Huang, Heyan and Wu, Zhijing},
    title = {Boosting legal case retrieval by query content selection with large language models},
    year = {2023},
    isbn = {9798400704086},
    publisher = {Association for Computing Machinery},
    address = {New York, NY, USA},
    url = {https://doi.org/10.1145/3624918.3625328},
    doi = {10.1145/3624918.3625328},
    booktitle = {Proceedings of the Annual International ACM SIGIR Conference on Research and Development in Information Retrieval in the Asia Pacific Region},
    pages = {176–184},
    numpages = {9},
    location = {Beijing, China},
    series = {SIGIR-AP '23}
}

@inproceedings{ma2023rewriteretrieveread,
    title = "Query Rewriting in Retrieval-Augmented Large Language Models",
    author = "Ma, Xinbei  and
      Gong, Yeyun  and
      He, Pengcheng  and
      Zhao, Hai  and
      Duan, Nan",
    editor = "Bouamor, Houda  and
      Pino, Juan  and
      Bali, Kalika",
    booktitle = "Proceedings of the 2023 Conference on Empirical Methods in Natural Language Processing",
    month = dec,
    year = "2023",
    address = "Singapore",
    publisher = "Association for Computational Linguistics",
    url = "https://aclanthology.org/2023.emnlp-main.322/",
    doi = "10.18653/v1/2023.emnlp-main.322",
    pages = "5303--5315"
}

@inproceedings{saleh2020documenttranslation,
    title = "Document Translation vs. Query Translation for Cross-Lingual Information Retrieval in the Medical Domain",
    author = "Saleh, Shadi  and
      Pecina, Pavel",
    editor = "Jurafsky, Dan  and
      Chai, Joyce  and
      Schluter, Natalie  and
      Tetreault, Joel",
    booktitle = "Proceedings of the 58th Annual Meeting of the Association for Computational Linguistics",
    month = jul,
    year = "2020",
    address = "Online",
    publisher = "Association for Computational Linguistics",
    url = "https://aclanthology.org/2020.acl-main.613/",
    doi = "10.18653/v1/2020.acl-main.613",
    pages = "6849--6860"
}

@article{zhang2023miracl,
    title = "{MIRACL}: A Multilingual Retrieval Dataset Covering 18 Diverse Languages",
    author = "Zhang, Xinyu  and
      Thakur, Nandan  and
      Ogundepo, Odunayo  and
      Kamalloo, Ehsan  and
      Alfonso-Hermelo, David  and
      Li, Xiaoguang  and
      Liu, Qun  and
      Rezagholizadeh, Mehdi  and
      Lin, Jimmy",
    journal = "Transactions of the Association for Computational Linguistics",
    volume = "11",
    year = "2023",
    address = "Cambridge, MA",
    publisher = "MIT Press",
    url = "https://aclanthology.org/2023.tacl-1.63/",
    doi = "10.1162/tacl_a_00595",
    pages = "1114--1131"
}

@incollection{rocchio1971relevance,
  author    = {Rocchio, Jr., Joseph J.},
  title     = {Relevance Feedback in Information Retrieval},
  editor    = {Salton, Gerard},
  booktitle = {The {SMART} Retrieval System: Experiments in Automatic Document Processing},
  chapter   = {14},
  pages     = {313--323},
  publisher = {Prentice-Hall},
  address   = {Englewood Cliffs, NJ},
  year      = {1971}
}

@article{wang2021colbertprf,
    author = {Wang, Xiao and MacDonald, Craig and Tonellotto, Nicola and Ounis, Iadh},
    title = {{ColBERT-PRF}: Semantic Pseudo-Relevance Feedback for Dense Passage and Document Retrieval},
    year = {2023},
    issue_date = {February 2023},
    publisher = {Association for Computing Machinery},
    address = {New York, NY, USA},
    volume = {17},
    number = {1},
    issn = {1559-1131},
    url = {https://doi.org/10.1145/3572405},
    doi = {10.1145/3572405},
    journal = {ACM Trans. Web},
    month = jan,
    articleno = {3},
    numpages = {39}
}

@misc{openai2025gptoss120bgptoss20bmodel,
      title={{gpt-oss-120b \& gpt-oss-20b Model Card}}, 
      author={{OpenAI} and Sandhini Agarwal and Lama Ahmad and Jason Ai and Sam Altman and Andy Applebaum and Edwin Arbus and Rahul K. Arora and Yu Bai and Bowen Baker and Haiming Bao and Boaz Barak and Ally Bennett and Tyler Bertao and Nivedita Brett and Eugene Brevdo and Greg Brockman and Sebastien Bubeck and Che Chang and Kai Chen and Mark Chen and Enoch Cheung and Aidan Clark and Dan Cook and Marat Dukhan and Casey Dvorak and Kevin Fives and Vlad Fomenko and Timur Garipov and Kristian Georgiev and Mia Glaese and Tarun Gogineni and Adam Goucher and Lukas Gross and Gil Guzman, Katia and John Hallman and Jackie Hehir and Johannes Heidecke and Alec Helyar and Haitang Hu and Romain Huet and Jacob Huh and Saachi Jain and Zach Johnson and Chris Koch and Irina Kofman and Dominik Kundel and Jason Kwon and Volodymyr Kyrylov and Elaine Ya Le and Guillaume Leclerc and James Park Lennon and Scott Lessans and Mario Lezcano-Casado and Yuanzhi Li and Zhuohan Li and Ji Lin and Jordan Liss and Liu, Lily (Xiaoxuan) and Jiancheng Liu and Kevin Lu and Chris Lu and Zoran Martinovic and Lindsay McCallum and Josh McGrath and Scott McKinney and Aidan McLaughlin and Song Mei and Steve Mostovoy and Tong Mu and Gideon Myles and Alexander Neitz and Alex Nichol and Jakub Pachocki and Alex Paino and Dana Palmie and Ashley Pantuliano and Giambattista Parascandolo and Jongsoo Park and Leher Pathak and Carolina Paz and Ludovic Peran and Dmitry Pimenov and Michelle Pokrass and Elizabeth Proehl and Huida Qiu and Gaby Raila and Filippo Raso and Hongyu Ren and Kimmy Richardson and David Robinson and Bob Rotsted and Hadi Salman and Suvansh Sanjeev and Max Schwarzer and D. Sculley and Harshit Sikchi and Kendal Simon and Karan Singhal and Yang Song and Dane Stuckey and Zhiqing Sun and Philippe Tillet and Sam Toizer and Foivos Tsimpourlas and Nikhil Vyas and Eric Wallace and Xin Wang and Miles Wang and Olivia Watkins and Kevin Weil and Amy Wendling and Kevin Whinnery and Cedric Whitney and Hannah Wong and Lin Yang and Yu Yang and Michihiro Yasunaga and Kristen Ying and Wojciech Zaremba and Wenting Zhan and Cyril Zhang and Brian Zhang and Eddie Zhang and Shengjia Zhao},
      year={2025},
      eprint={2508.10925},
      archivePrefix={arXiv},
      primaryClass={cs.CL},
      doi={10.48550/arXiv.2508.10925},
      url={https://arxiv.org/abs/2508.10925}, 
}

@inproceedings{smucker2007comparison,
    author = {Smucker, Mark D. and Allan, James and Carterette, Ben},
    title = {A comparison of statistical significance tests for information retrieval evaluation},
    year = {2007},
    isbn = {9781595938039},
    publisher = {Association for Computing Machinery},
    address = {New York, NY, USA},
    url = {https://doi.org/10.1145/1321440.1321528},
    doi = {10.1145/1321440.1321528},
    booktitle = {Proceedings of the Sixteenth ACM Conference on Conference on Information and Knowledge Management},
    pages = {623–632},
    numpages = {10},
    location = {Lisbon, Portugal},
    series = {CIKM '07}
}

@article{holm1979simple,
  author  = {Holm, Sture},
  title   = {A Simple Sequentially Rejective Multiple Test Procedure},
  journal = {Scandinavian Journal of Statistics},
  volume  = {6},
  number  = {2},
  pages   = {65--70},
  year    = {1979},
  doi     = {10.2307/4615733},
  url     = {https://www.jstor.org/stable/4615733}
}

\appendix
\section*{Appendix}
\section{Legal Area Statistics}
\label{app:legal_area_statistics}

Table~\ref{tab:gbb_area_stats} reports GreekBarRetrieval statistics by legal area. Article counts are multi-label because some statutory sources are used in more than one GreekBarBench legal area.

\begin{table}[!ht]
  \centering
  \small
  \setlength{\tabcolsep}{4pt}
  \begin{tabular}{lcrrcc}
    \toprule
    \textbf{Legal area} & \textbf{Articles} & \textbf{Q} & \textbf{Qrels} & \textbf{Rel./Q} & \textbf{MedLen} \\
    \midrule
    Civil      & 3{,}264 & 69 & 255 & 3.70 & 42 \\
    Public     & 2{,}841 & 58 & 99  & 1.71 & 41 \\
    Commercial & 4{,}171 & 48 & 127 & 2.65 & 48 \\
    Lawyers    & 4{,}476 & 56 & 153 & 2.73 & 48 \\
    Criminal   & 1{,}253 & 52 & 141 & 2.71 & 85 \\
    \bottomrule
  \end{tabular}
  \caption{\textbf{Legal area statistics} for GreekBarRetrieval. Article counts are multi-label and, therefore, exceed the 6{,}308 unique articles in the full retrieval pool. Q denotes the number of queries, Qrels the total number of gold articles (for all queries of the area together), Rel./Q the mean number of relevant articles per query. MedLen is the median article length in words.}
  \vspace{-4mm}
  \label{tab:gbb_area_stats}
\end{table}

\section{Legal Source (Prefix) to Area Mapping}
\label{app:prefix_area_mapping}

GreekBarRetrieval uses article identifiers of the form \texttt{PREFIX::ARTICLE}. The prefix denotes the legal source the article is drawn from. Table~\ref{tab:prefix_domain_mapping} reports the mapping used for area-level analysis. The original GreekBarBench legal area tags are \texttt{astiko} (Αστικό Δίκαιο, Civil Law), \texttt{dimosio} (Δημόσιο Δίκαιο, Public Law), \texttt{emporiko} (Εμπορικό Δίκαιο, Commercial Law), \texttt{kodikas} (Κώδικας Δικηγόρων, Lawyers' Code), and \texttt{poiniko} (Ποινικό Δίκαιο, Criminal Law).

\begin{table*}[t]
\centering
\small
\setlength{\tabcolsep}{5pt}
\renewcommand{\arraystretch}{1.05}
\begin{tabular}{@{}l>{\raggedright\arraybackslash}p{0.36\textwidth}>{\raggedright\arraybackslash}p{0.38\textwidth}@{}}
\toprule
\textbf{Prefix} & \textbf{Source description} & \textbf{Legal Area} \\
\midrule
\texttt{AK} & Greek Civil Code & Civil, Public, Commercial Law, Lawyers' Code \\
\texttt{EisNAK} & Introductory Law to the Civil Code & Public Law \\
\texttt{KD} & Lawyers' Code & Lawyers' Code \\
\texttt{KDD} & Code of Administrative Procedure & Public Law \\
\texttt{KDD/sias} & Source prefix retained from GreekBarBench & Public Law \\
\texttt{KDDL} & Source prefix retained from GreekBarBench & Lawyers' Code \\
\texttt{KPD} & Code of Criminal Procedure & Lawyers' Code, Criminal Law \\
\texttt{KPolD} & Code of Civil Procedure & Civil Law, Commercial Law, Lawyers' Code \\
\texttt{PK} & Greek Penal Code & Lawyers' Code, Criminal Law \\
\texttt{Syntagma} & Constitution of Greece & Civil Law, Public Law, Criminal Law \\
\texttt{N\_702\_1977} & Law 702/1977 & Public Law \\
\texttt{N\_1406\_1983} & Law 1406/1983 & Public Law \\
\texttt{N\_3155\_1955} & Law 3155/1955 & Public Law \\
\texttt{PD\_18\_1989} & Presidential Decree 18/1989 & Public Law \\
\texttt{PD\_258\_2005} & Presidential Decree 258/2005 & Public Law \\
\texttt{N\_146\_1914} & Law 146/1914 & Commercial Law \\
\texttt{N\_3190\_1955} & Law 3190/1955 & Commercial Law \\
\texttt{N\_4072\_2012} & Law 4072/2012 & Commercial Law \\
\texttt{N\_4541\_2018} & Law 4541/2018 & Commercial Law \\
\texttt{N\_4738\_2020} & Law 4738/2020 & Commercial Law \\
\texttt{N\_5325\_1932} & Law 5325/1932 & Commercial Law \\
\texttt{N\_5960\_1933} & Law 5960/1933 & Commercial Law \\
\texttt{ND\_17\_07\_N\_13\_08\_1923} & Legislative decree of 17-07/13-08-1923 & Commercial Law \\
\bottomrule
\end{tabular}
\caption{\textbf{Mapping from legal sources (prefixes) to GreekBarBench legal areas}. The mapping is multi-label, because some statutory sources are used in more than one legal area.}
\vspace{-3mm}
\label{tab:prefix_domain_mapping}
\end{table*}


\section{Detailed Results}
\label{app:detailed_results}
Table~\ref{tab:main_results} reported experimental results for representative systems from each experimental setting. This appendix provides additional results for all evaluated retrieval methods and configurations.

\subsection{Detailed Results of Vanilla Retrievers}
\label{app:results_oneshot_sparse_dense}

Table~\ref{tab:app_oneshot_sparse_dense} reports the experimental results of all vanilla retrievers tested. The results of the dense retrievers are as in Table~\ref{tab:main_results}, but the sparse results now include all three BM25 variants tested (\S\ref{subsec:retrieval_baselines}, Appendix~\ref{app:retriever_details}). 
Within the Qwen3 family, performance increases with model parameter count. \texttt{Nomic-v1.5} is a clear outlier, with 0.08 Recall@100, possibly because it was trained only on English data. \textsc{BM25-GreekStemmer} and \textsc{BM25-spaCy} obtain the same results (rounded to two decimals). \textsc{BM25-gr-nlp} is clearly worse.

\begin{table}[!ht]
\centering
\footnotesize
\setlength{\tabcolsep}{3pt}
\renewcommand{\arraystretch}{1.02}
\begin{tabular}{@{}>{\raggedright\arraybackslash}p{0.40\linewidth}cccc@{}}
\toprule
\textbf{Model} & \textbf{N@10} & \textbf{N@100} & \textbf{R@10} & \textbf{R@100} \\
\midrule

\rowcolor{headerbg}
\multicolumn{5}{l}{\textit{Dense retrieval, no query reformulation}} \\
\texttt{Gemini-001}
& \textbf{0.39} & \textbf{0.47} & \textbf{0.48} & \textbf{0.77} \\
\texttt{Qwen3-8B}
& \localbest{0.28} & \localbest{0.36} & \localbest{0.38} & 0.67 \\
\texttt{Euler-Legal-V1}
& 0.25 & 0.34 & 0.34 & \localbest{0.68} \\
\texttt{Qwen3-4B}
& 0.24 & 0.32 & 0.31 & 0.62 \\
\texttt{Jina-v5-small}
& 0.22 & 0.29 & 0.29 & 0.57 \\
\texttt{Qwen3-0.6B}
& 0.14 & 0.20 & 0.20 & 0.45 \\
\texttt{Arctic-v2}
& 0.22 & 0.28 & 0.28 & 0.51 \\
\texttt{EmbGemma-300M}
& 0.10 & 0.16 & 0.15 & 0.38 \\
\texttt{Nomic-v1.5}
& 0.01 & 0.02 & 0.01 & 0.08 \\

\midrule
\rowcolor{headerbg}
\multicolumn{5}{l}{\textit{Sparse retrieval, no query reformulation}} \\
\textsc{BM25-GreekStemmer}
& 0.10 & 0.14 & 0.16 & 0.36 \\
\textsc{BM25-spaCy}
& 0.10 & 0.14 & 0.16 & 0.36 \\
\textsc{BM25-gr-nlp}
& 0.05 & 0.07 & 0.08 & 0.18 \\

\bottomrule
\end{tabular}
\caption{\textbf{Detailed vanilla sparse and dense retrieval} results. The results of the dense retrievers are the same as in Table~\ref{tab:main_results}.  
\textsc{BM25-GreekStemmer} and \textsc{BM25-spaCy} are the two best BM25 variants and are effectively indistinguishable (their results differ in the third decimal, but results are rounded to two decimals here). 
Bold marks the best overall score. Blue shading marks the best score among locally hosted systems.}
\vspace{-6mm}
\label{tab:app_oneshot_sparse_dense}
\end{table}

\subsection{Results for Translation, Query Reformulation, PRF, and RRF}
\label{app:results_interventions}

Table~\ref{tab:app_interventions} shows the effect of English translation, query reformulation, pseudo-relevance feedback (PRF), and hybrid sparse-dense retrieval (RRF) on each vanilla retriever, reporting Recall@100.
 \texttt{Gemini-001} is omitted because we ran it only on the original Greek queries. We also omitted some of the experiments (dashes), when the other results (in the same columns) were not promising.  

\begin{table*}[!t]
\centering
\small
\setlength{\tabcolsep}{6pt}
\renewcommand{\arraystretch}{1.07}
\begin{tabular}{@{}>{\raggedright\arraybackslash}p{0.26\textwidth}ccccc@{}}
\toprule
\textbf{Retriever} & \textbf{Vanilla} & \textbf{+Translation} & \textbf{+Reformulation} & \textbf{+PRF} & \textbf{+RRF} \\
\midrule
\texttt{Qwen3-8B} & 0.67 & 0.71 ($+0.04$) & \best{0.73} ($+0.06$) & 0.66 ($-0.01$) & 0.64 ($-0.03$) \\
\texttt{Euler-Legal-V1} & 0.68 & 0.68 ($+0.00$) & 0.71 ($+0.03$) & --- & 0.63 ($-0.05$) \\
\texttt{Qwen3-4B} & 0.62 & 0.67 ($+0.05$) & 0.69 ($+0.07$) & --- & 0.58 ($-0.04$) \\
\texttt{Jina-v5-small} & 0.57 & 0.65 ($+0.08$) & 0.63 ($+0.06$) & --- & 0.56 ($-0.01$) \\
\texttt{Arctic-v2} & 0.51 & 0.55 ($+0.04$) & 0.60 ($+0.09$) & --- & 0.50 ($-0.01$) \\
\texttt{Qwen3-0.6B} & 0.45 & 0.55 ($+0.10$) & 0.48 ($+0.03$) & --- & 0.46 ($+0.01$) \\
\texttt{EmbGemma-300M} & 0.38 & 0.56 ($+0.18$) & 0.41 ($+0.03$) & 0.36 ($-0.02$) & 0.42 ($+0.04$) \\
\texttt{Nomic-v1.5} & 0.08 & 0.42 ($+0.34$) & 0.07 ($-0.01$) & --- & 0.32 ($+0.24$) \\
\midrule
\textsc{BM25-GreekStemmer} & 0.36 & 0.25 ($-0.11$) & \best{0.60} ($+0.23$) & 0.37 ($+0.01$) & --- \\
\textsc{BM25-spaCy} & 0.36 & 0.38 ($+0.02$) & 0.56 ($+0.20$) & 0.37 ($+0.01$) & --- \\
\textsc{BM25-GR-NLP} & 0.18 & 0.24 ($+0.06$) & 0.48 ($+0.30$) & 0.19 ($+0.01$) & --- \\
\bottomrule
\end{tabular}
\caption{\textbf{Detailed results (Recall@100)} showing the effect of separately adding \textbf{English translation}, \textbf{LLM query reformulation}, \textbf{pseudo-relevance feedback (PRF)}, and \textbf{sparse-dense fusion (RRF)} to the vanilla retrievers. Parentheses report the 
difference from the corresponding vanilla result. Dashes denote combinations we did not run, given that the other experiments in the column were not promising.}
\vspace{-4mm}
\label{tab:app_interventions}
\end{table*}

Reformulation is the only addition (among the four) that helps both dense and sparse retrievers, and the only one whose gain is consistent and large anywhere. It adds 0.23 to \textsc{BM25-GreekStemmer} and 0.20 to \textsc{BM25-spaCy}, against 0.03 to 0.09 for the dense encoders that work on Greek. Translation shows the opposite pattern within the dense encoders, helping the weakest most, and having a negative effect ($-0.11$) on \textsc{BM25-GreekStemmer}. Across the five systems evaluated with PRF, the absolute change in Recall@100 is at most $0.02$: PRF improves each of the three sparse retrievers by $+0.01$, but decreases \texttt{Qwen3-8B} by $-0.01$ and \texttt{EmbGemma-300M} by $-0.02$. Fusion deteriorates the performance of five of the eight dense encoders. There is a large gain ($+0.24$) only for \texttt{Nomic-v1.5}, a dense model not performing well on Greek, hence BM25's exact matching helps.

\subsection{Query Reformulation Example}
\label{app:qualitative_reformulation_example}

\begin{table}[!t]
\centering
\footnotesize
\setlength{\tabcolsep}{0pt}
\renewcommand{\arraystretch}{1.05}
\begin{tabular}{p{0.99\linewidth}}
\toprule

\textbf{Original Query (abridged)} \\
\midrule
Has A committed an offence, and if so which one? A, a municipal
employee responsible for receiving applications for a position in a
municipal enterprise, unlawfully refused to accept B's documents so
that A's relative would be hired. \\

\midrule
\textbf{Reformulated Keyword Query (abridged)} \\
\midrule
municipal employee; responsibility for receiving application
documents; refusal to accept documents; hiring a relative;
\textbf{breach of duty}; criminal liability of a public employee;
\textbf{unlawful benefit}; personal preference \\

\midrule
\textbf{Gold Article: \texttt{CrimC::259} (excerpt)} \\
\midrule
An employee who intentionally \textbf{breaches the duties of their
office} with the purpose of obtaining for themselves or another an
\textbf{unlawful benefit} is punishable by imprisonment. \\

\bottomrule
\end{tabular}
\vspace{-2mm}
\caption{\textbf{Query reformulation example}.
The original query (question+facts) is converted (for sparse retrieval) to a legal keyword query. Bold
highlights terminology shared by the reformulation and the gold article. Example translated to
English and abridged for presentation.}
\vspace{-4mm}
\label{tab:qualitative_reformulation_example}
\end{table}

In the original query of Table~\ref{tab:qualitative_reformulation_example}, the gold article is outside the top 100 of 
\textsc{BM25-GreekStemmer}. Query  reformulation introduces the legal term \emph{παράβαση καθήκοντος} (breach of duty),  matching (after stemming) \emph{παραβαίνει τα καθήκοντα} (breaches the duties) in the statutory text, and moves the article to rank 18. 
This example illustrates how reformulation provides lexical cues that particularly benefit sparse retrieval.

\subsection{\textsc{ReAct-BM25} Results}
\label{app:results_agentic_bm25}

Table~\ref{tab:app_agentic_bm25_metrics} shows that all five metrics improve at every reported round of \textsc{ReAct-BM25}. Recall@100 increases from 0.45 after round 1 to 0.67 after round 10, while most of the improvement in the ranking metrics occurs during the first two rounds.

\begin{table}[t]
\centering
\small
\setlength{\tabcolsep}{3pt}
\renewcommand{\arraystretch}{1.02}
\begin{tabular}{@{}cccccc@{}}
\toprule
\textbf{Rounds} & \textbf{N@10} & \textbf{N@100} & \textbf{R@10} & \textbf{R@100} & \textbf{MAP} \\
\midrule
0  & 0.10 & 0.14 & 0.16 & 0.36 & 0.09 \\
1  & 0.34 & 0.35 & 0.44 & 0.45 & 0.27 \\
2  & 0.38 & 0.39 & 0.47 & 0.52  & 0.30 \\
3  & 0.40 & 0.42 & 0.49 & 0.55  & 0.33 \\
5  & 0.41 & 0.45 & 0.50 & 0.62  & 0.35 \\
10 & \best{0.43} & \best{0.47} & \best{0.52} & \best{0.67} & \best{0.37} \\
\bottomrule
\end{tabular}
\caption{\textbf{Detailed \textsc{ReAct-BM25} results}. Round 0 is the vanilla \textsc{BM25-GreekStemmer}. 
}
\vspace{-5mm}
\label{tab:app_agentic_bm25_metrics}
\end{table}

To separate candidate article generation (articles retrieved by the planner) from observer filtering, we compute recall over the union of all unique BM25 candidates retrieved up to each round (\emph{candidate pool recall}), before the observer's decisions are applied. Table~\ref{tab:react_candidate_recall} shows that candidate-pool recall increases from 0.59 after the first round to 0.77 after ten rounds, compared with 0.45 and 0.67, respectively, for the observer-kept output. By round 10, 98 retrieved (by the planner) relevant query-article pairs (an article may be relevant to many queries), affecting 64 queries, are absent from the observer-kept output. Thus, candidate generation retrieves more relevant evidence than the final output preserves, showing that observer filtering accounts for part of the low (0.67) Recall@100 (Table~\ref{tab:main_results}).

\begin{table*}[t]
\centering
\footnotesize
\setlength{\tabcolsep}{7pt}
\renewcommand{\arraystretch}{1.08}

\begin{tabular}{@{}rcccrr@{}}
\toprule
\textbf{Round} &
\textbf{Observer-kept R@100} &
\textbf{Candidate-pool R@100} &
\textbf{Gap} &
\textbf{Discarded gold pairs} &
\textbf{Affected queries} \\
\midrule
1  & 0.45 & 0.59 & 0.14 & 125 & 86 \\
2  & 0.52 & 0.67 & 0.15 & 130 & 87 \\
3  & 0.56 & 0.72 & 0.16 & 131 & 91 \\
5  & 0.63 & 0.74 & 0.12 & 111 & 74 \\
10 & 0.67 & 0.77 & 0.10 & 98  & 64 \\
\bottomrule
\end{tabular}

\caption{\textbf{Effect of observer filtering on Recall@100 in \textsc{ReAct-BM25}.}
Candidate-pool recall R@100 is computed over the union of all unique BM25
candidates retrieved up to each round, ignoring the filtering decisions of the observer. Observer-kept R@100 includes the filtering of the observers. Gap is the difference between candidate-pool
Recall@100 and observer-kept Recall@100. Discarded
gold pairs are relevant query-article pairs (an article may be relevant to multiple queries) retrieved by BM25 but
excluded from the observer-kept output. 
}
\label{tab:react_candidate_recall}
\end{table*}

\subsection{Statistical Significance Tests}
\label{app:significance_tests}

Table~\ref{tab:significance_tests} reports two-sided paired randomization statistical significance tests~\citep{smucker2007comparison} for the main comparisons discussed in the paper. Rather than testing every possible system pair, we focus on the effect of query reformulation (Reform-) on \textsc{BM25} and \texttt{Qwen3-8B}, the additional improvement obtained by the iterative loop (\textsc{ReACT}-) over Reform-\textsc{BM25}, and the comparison of \textsc{ReAct-BM25} with Reform-\texttt{Qwen3-8B} and \texttt{Gemini-001}. To save resources and space, we consider nDCG@10, nDCG@100, Recall@100, and MAP@100, ignoring Recall@10 here.

All systems are evaluated on the same 283 queries. For each comparison and metric, we compute the mean score difference between the two systems over all queries. We then randomly swap the two system labels independently for each query and recompute the mean difference. We repeat this process 100K times and use a two-sided test, counting randomized differences at least as large as the observed difference in absolute value.
Since we perform 20 tests, we apply Holm correction to the resulting $p$ values~\citep{holm1979simple}.

\newcommand{\nostar}{\phantom{{}^{*}}}

\begin{table*}[t]
\centering
\footnotesize
\setlength{\tabcolsep}{3.5pt}
\renewcommand{\arraystretch}{1.08}

\begin{tabular}{@{}
>{\raggedright\arraybackslash}p{0.27\textwidth}
rrrrrrrr
@{}}
\toprule
\textbf{Comparison ($A-B$)}
& \multicolumn{2}{c}{\textbf{N@10}}
& \multicolumn{2}{c}{\textbf{N@100}}
& \multicolumn{2}{c}{\textbf{R@100}}
& \multicolumn{2}{c}{\textbf{MAP@100}} \\
\cmidrule(lr){2-3}
\cmidrule(lr){4-5}
\cmidrule(lr){6-7}
\cmidrule(l){8-9}
& \textbf{$\Delta$} & \textbf{$p_{\mathrm{H}}$}
& \textbf{$\Delta$} & \textbf{$p_{\mathrm{H}}$}
& \textbf{$\Delta$} & \textbf{$p_{\mathrm{H}}$}
& \textbf{$\Delta$} & \textbf{$p_{\mathrm{H}}$} \\
\midrule

\multicolumn{9}{@{}l}{\textit{Query reformulation}} \\

Reform-\textsc{BM25} vs.\ \textsc{BM25}
& $+0.10$ & $<.001^{*}$
& $+0.12$ & $<.001^{*}$
& $+0.23$ & $<.001^{*}$
& $+0.07$ & $<.001^{*}$ \\

Reform-\texttt{Qwen3-8B} vs.\ \texttt{Qwen3-8B}
& $+0.05$ & $.002^{*}$
& $+0.05$ & $<.001^{*}$
& $+0.06$ & $.005^{*}$
& $+0.04$ & $.007^{*}$ \\
\midrule
\multicolumn{9}{@{}l}{\textit{Iterative retrieval}} \\

\textsc{ReAct-BM25} vs.\ Reform-\textsc{BM25}
& $+0.23$ & $<.001^{*}$
& $+0.20$ & $<.001^{*}$
& $+0.08$ & $.002^{*}$
& $+0.21$ & $<.001^{*}$ \\

\textsc{ReAct-BM25} vs.\ Reform-\texttt{Qwen3-8B}
& $+0.10$ & $<.001^{*}$
& $+0.06$ & $.005^{*}$
& $-0.05$ & $.084\nostar$
& $+0.09$ & $<.001^{*}$ \\

\textsc{ReAct-BM25} vs.\ \texttt{Gemini-001}
& $+0.04$ & $.108\nostar$
& $+0.00$ & $.916\nostar$
& $-0.10$ & $<.001^{*}$
& $+0.03$ & $.156\nostar$ \\

\bottomrule
\end{tabular}

\caption{\textbf{Statistical significance tests} for the main system
comparisons. $\Delta$ is the difference between the mean scores of
systems $A$ (mentioned first in each comparison) and $B$ (mentioned second); positive values favor system $A$.
$p_{\mathrm{H}}$ is the Holm-adjusted $p$-value across the 20 tests.
An asterisk marks statistical significance at
$p_{\mathrm{H}}<0.05$.}
\label{tab:significance_tests}
\end{table*}

Table~\ref{tab:significance_tests} shows that query reformulation significantly improves \textsc{BM25} and \texttt{Qwen3-8B} on all four metrics. \textsc{ReAct-BM25} also significantly improves over Reform-\textsc{BM25} in all metrics. Compared with Reform-\texttt{Qwen3-8B}, \textsc{ReAct-BM25} obtains significantly higher nDCG@10, nDCG@100, MAP@100, but the Recall@100 difference is not significant.
Compared with \texttt{Gemini-001}, \textsc{ReAct-BM25} has significantly lower Recall@100, while the differences in nDCG@10, nDCG@100, and MAP are not statistically significant by our test.

\section{BM25 Parameter Tuning}
\label{app:bm25_tuning}

All BM25 systems in the main text use the \texttt{rank\_bm25} defaults (Table~\ref{tab:app_sparse_retrievers}), $k_1=1.5$,  $b=0.75$. As a sensitivity analysis, we swept $k_1 \in [0.2, 3.0]$ in steps of
approx.\ 0.2, and $b \in [0, 1]$ in steps of 0.1, also including the default values, for \textsc{BM25-GreekStemmer} on both the original and reformulated
queries, selecting the parameter values with the highest Recall@100.  Table~\ref{tab:app_bm25_tuning} reports the original and tuned results. Tuned parameter values were selected on the full benchmark and, therefore, the tuned results are provided only as oracle upper bounds (see also the Limitations). 

Parameter tuning adds 0.05 to Recall@100 when using the original queries, but only 0.01 when queries are reformulated. By contrast, adding query reformulation to vanilla BM25 adds 0.23 to Recall@100. With reformulated queries, tuning for Recall@100 leads to inferior  nDCG@10, nDCG@100, Recall@10 and MAP@100 scores.

\begin{table}[t]
\centering
\footnotesize
\setlength{\tabcolsep}{2.5pt}
\renewcommand{\arraystretch}{0.97}
\begin{tabular}{@{}>{\raggedright\arraybackslash}p{0.30\linewidth}ccccc@{}}
\toprule
\textbf{System} & \textbf{N@10} & \textbf{N@100} & \textbf{R@10} & \textbf{R@100} & \textbf{MAP} \\
\midrule
\rowcolor{headerbg}
\multicolumn{6}{@{}l}{\textit{Original queries}} \\
\textsc{BM25}        & 0.10 & 0.14 & 0.16 & 0.36 & 0.09 \\
\quad tuned          & 0.10 & 0.16 & 0.16 & 0.41 & 0.07 \\
\midrule
\rowcolor{headerbg}
\multicolumn{6}{@{}l}{\textit{Reformulated queries}} \\
Reform-\textsc{BM25} & 0.20 & 0.28 & 0.30 & 0.60 & 0.16 \\
\quad tuned          & 0.18 & 0.26 & 0.25 & 0.61 & 0.14 \\
\bottomrule
\end{tabular}
\caption{\textbf{Effect of tuning parameters} $k_1$ and $b$ for \textsc{BM25-GreekStemmer}, with and without LLM query reformulation. Tuned values are selected on the full benchmark and, therefore, the tuned results are oracle upper bounds. Metric abbreviations follow Table~\ref{tab:main_results}.}
\vspace{-4mm}
\label{tab:app_bm25_tuning}
\end{table}

\section{Retriever Implementation Details}
\label{app:retriever_details}

Tables~\ref{tab:app_sparse_retrievers}--\ref{tab:app_dense_retrievers} give the full identifier for every retriever, together with the settings we ran it with. All sparse systems use the \texttt{rank\_bm25} implementation with default parameters; Appendix~\ref{app:bm25_tuning} reports what tuning changes. All dense systems use the released checkpoint with no fine-tuning on Greek legal text, and encode each article as a single passage.

\begin{table*}[t]
\centering
\small
\setlength{\tabcolsep}{4pt}
\renewcommand{\arraystretch}{1.10}

\begin{tabular}{@{}
  >{\raggedright\arraybackslash}p{0.20\textwidth}
  >{\raggedright\arraybackslash}p{0.18\textwidth}
  >{\raggedright\arraybackslash}p{0.40\textwidth}
  >{\raggedright\arraybackslash}p{0.12\textwidth}
@{}}
\toprule
\textbf{BM25 variant} &
\textbf{Resource ID} &
\textbf{Implementation details} &
\textbf{Resource} \\
\midrule

BM25 base
& \texttt{rank\_bm25}
& The default sparse baselines use \texttt{BM25Okapi} with
$k_1=1.5$ and $b=0.75$. The parameter-tuning experiments in
Table~\ref{tab:app_bm25_tuning} vary these parameters only for
sensitivity analysis.
& \href{https://github.com/dorianbrown/rank_bm25}
        {GitHub} \\

\textsc{BM25-GreekStemmer}
& \texttt{greek-stemmer}
& Uses Unicode-aware regular-expression tokenization with
\texttt{\textbackslash w+}, followed by diacritic stripping,
uppercasing, removal of a manually defined Greek stopword list,
and stemming with \texttt{GreekStemmer}.
& \href{https://github.com/alup/python_greek_stemmer}
        {GitHub} \\

\textsc{BM25-spaCy}
& \texttt{el\_core\_news\_sm}
& Uses the spaCy Greek pipeline. Whitespace and punctuation tokens
are removed; lemmas are used when available; and tokens are
lowercased and stripped of diacritics. The pipeline also removes
spaCy's Greek stopwords.
& \href{https://spacy.io/models/el}
        {spaCy docs} \\

\textsc{BM25-gr-nlp-toolkit}
& \texttt{gr-nlp-toolkit}
& Uses the toolkit tokenizer and processes long texts in chunks.
Punctuation tokens are removed, but no stemming, lemmatization,
lowercasing, or diacritic stripping is applied.
& \href{https://github.com/nlpaueb/gr-nlp-toolkit}
        {GitHub} \\

\textsc{BM25-English-spaCy}
& \texttt{en\_core\_web\_sm}
& Uses the spaCy English pipeline for the English translation
baseline. Whitespace and punctuation tokens are removed; lemmas
are used when available; and tokens are lowercased. The pipeline
also removes spaCy's English stopwords.
& \href{https://spacy.io/models/en}
        {spaCy docs} \\

\bottomrule
\end{tabular}

\caption{\textbf{Sparse retrieval implementation resources}. The three Greek
BM25 variants differ in their preprocessing, while
\textsc{BM25-English-spaCy} is used only for the English translation
baseline. Unless otherwise indicated, all systems use the default
\texttt{BM25Okapi} parameters.}
\label{tab:app_sparse_retrievers}
\end{table*}

\begin{table*}[t]
\centering
\small
\setlength{\tabcolsep}{3.5pt}
\renewcommand{\arraystretch}{1.10}

\begin{tabular}{@{}
  >{\raggedright\arraybackslash}p{0.21\textwidth}
  >{\raggedright\arraybackslash}p{0.14\textwidth}
  >{\raggedright\arraybackslash}p{0.38\textwidth}
  >{\raggedright\arraybackslash}p{0.17\textwidth}
@{}}
\toprule
\textbf{Retriever} &
\textbf{Interface} &
\textbf{Implementation details} &
\textbf{Resource} \\
\midrule

\texttt{Gemini-001}
& Google GenAI API
& API-based encoder. The output dimensionality is set to 3072.
retrieval pool passages are embedded with the retrieval-document task type,
and queries with the retrieval-query task type.
& \href{https://ai.google.dev/gemini-api/docs/models/gemini-embedding-001}
        {Google documentation} \\

\texttt{Qwen3-8B}
& LM Studio
& Local encoder served through the LM Studio HTTP embeddings endpoint.
Queries are prefixed with a legal-retrieval instruction using the Qwen3
instruction format; documents are embedded as raw text.
& \href{https://huggingface.co/Qwen/Qwen3-Embedding-8B}
        {Model card} \\

\texttt{Qwen3-4B}
& LM Studio
& Local encoder served through the LM Studio HTTP embeddings endpoint.
It uses the same Qwen3 instruction-aware query formatting as the 8B
variant; documents are embedded as raw text.
& \href{https://huggingface.co/Qwen/Qwen3-Embedding-4B}
        {Model card} \\

\texttt{Qwen3-0.6B}
& LM Studio
& Local encoder served through the LM Studio HTTP embeddings endpoint.
It uses the same instruction-aware query formatting as the larger Qwen3
variants; documents are embedded as raw text.
& \href{https://huggingface.co/Qwen/Qwen3-Embedding-0.6B}
        {Model card} \\

\texttt{Euler-Legal-V1}
& Local HF
& Legal-domain encoder served through a custom local Hugging Face HTTP
embeddings endpoint. Queries and documents are embedded without
additional prefixes.
& \href{https://huggingface.co/Mira190/Euler-Legal-Embedding-V1}
        {Model card} \\

\texttt{Jina-v5-small}
& Local HF
& Multilingual retrieval encoder served through a custom local Hugging
Face HTTP embeddings endpoint. Queries use a \texttt{Query:} prefix,
and documents use a \texttt{Document:} prefix.
& \href{https://huggingface.co/jinaai/jina-embeddings-v5-text-small}
        {Model card} \\

\texttt{Arctic-v2}
& Local HF
& Multilingual retrieval encoder served through a custom local Hugging
Face HTTP embeddings endpoint. Queries are prefixed with
\texttt{query:}; documents are embedded as raw text.
& \href{https://huggingface.co/Snowflake/snowflake-arctic-embed-l-v2.0}
        {Model card} \\

\texttt{EmbGemma-300M}
& LM Studio
& Local encoder served through the LM Studio HTTP embeddings endpoint.
Queries use the \texttt{task: search result | query:} format, and
documents use the \texttt{title: none | text:} format.
& \href{https://huggingface.co/google/embeddinggemma-300m}
        {Model card} \\

\texttt{Nomic-v1.5}
& LM Studio
& Local encoder served through the LM Studio HTTP embeddings endpoint.
Queries use the \texttt{search\_query:} prefix, and documents use the
\texttt{search\_document:} prefix.
& \href{https://huggingface.co/nomic-ai/nomic-embed-text-v1.5}
        {Model card} \\

\bottomrule
\end{tabular}

\caption{\textbf{Dense retrieval implementation resources}. With the exception
of \texttt{Gemini-001}, which is accessed through the Google GenAI
API, all dense retrievers are served locally through HTTP embedding
endpoints: LM Studio for the Qwen3, EmbGemma-300M, and Nomic models, and a
custom Hugging Face endpoint for Snowflake-Arctic, Jina, and
Euler-Legal. All dense runs use L2-normalized embeddings and exact
top-$k$ retrieval.}
\label{tab:app_dense_retrievers}
\end{table*}

\section{Further Inference Cost Analysis}
\label{app:cost_analysis}

Table~\ref{tab:cost_analysis} reports what each system costs to run one query. We separate embedding tokens from LLM tokens because they are priced differently, and we report retrieval pool embedding as a one-time cost rather than a per-query one, since it is paid once and reused. LLM token counts include input and output. For \textsc{ReAct-BM25} the input prevails: the observer prompt includes the retrieved article text, so most of the ten-round total belongs to the prompt rather than the generation. Dollar figures are estimates at current API rates and are given for comparison between systems, not as the cost we paid.

\begin{table*}[t]
\centering
\footnotesize
\setlength{\tabcolsep}{3.5pt}
\renewcommand{\arraystretch}{1.08}
\begin{tabular}{>{\raggedright\arraybackslash}p{0.30\textwidth}rrrrrr}
\toprule
\textbf{System} &
\textbf{Calls/q} &
\textbf{Time/q} &
\textbf{Embed tok/q} &
\textbf{LLM tok/q} &
\textbf{USD/q} &
\textbf{retrieval pool cost} \\
\midrule

\multicolumn{7}{l}{\textit{Vanilla baselines}} \\

\texttt{Qwen3-8B}
& 1 & 1.23s & 1.5k & -- & 0.000015 & 0.0385 \\

\texttt{Qwen3-4B}
& 1 & 0.78s & 1.5k & -- & 0.000031 & 0.0764 \\

\textsc{BM25-GreekStemmer}
& 0 & 0.30s & -- & -- & 0.000000 & -- \\

\midrule
\multicolumn{7}{l}{\textit{LLM query reformulation}} \\

Reform-\texttt{Qwen3-8B}
& 2 & 5.13s & 0.3k & 2.6k & 0.000630 & 0.0385 \\

Reform-\textsc{BM25-GreekStemmer}
& 1 & 0.06s & -- & 2.9k & 0.000613 & -- \\

\midrule
\multicolumn{7}{l}{\textit{Agentic retrieval}} \\

\textsc{ReAct-BM25} (10 rounds)
& 20 & 26.6m & -- & 617.9k & 0.096802 & -- \\

\bottomrule
\end{tabular}

\caption{\textbf{Inference cost analysis for representative retrieval systems}.
\emph{Calls/q} counts online model invocations per query;
\emph{Time/q} reports measured runtime per query, including reformulation;
\emph{Embed tok/q} reports embedding input tokens per query;
\emph{LLM tok/q} reports LLM input and output tokens per query;
\emph{USD/q} reports estimated equivalent API cost in US dollars;
and \emph{retrieval pool cost} reports the estimated one-time cost, in US dollars with current rates, of embedding the full retrieval pool.}

\label{tab:cost_analysis}
\end{table*}

\section{Prompts}
\label{app:prompts}

This appendix presents English translations of all prompt templates used in our experiments. The original prompts, inputs, and outputs were in Greek, except for the translation baseline, whose outputs were in English. We preserve the structure and content of the original prompts as faithfully as possible to support reproducibility. The prompts were not tuned using the benchmark results. The actual Greek prompts are included in our code repository.

\subsection{Translation and Reformulation}
\label{app:translation_reformulation}

\begin{figure*}[p]
    \centering
        \includegraphics[
        width=\textwidth,
        height=0.78\textheight,
        keepaspectratio
    ]{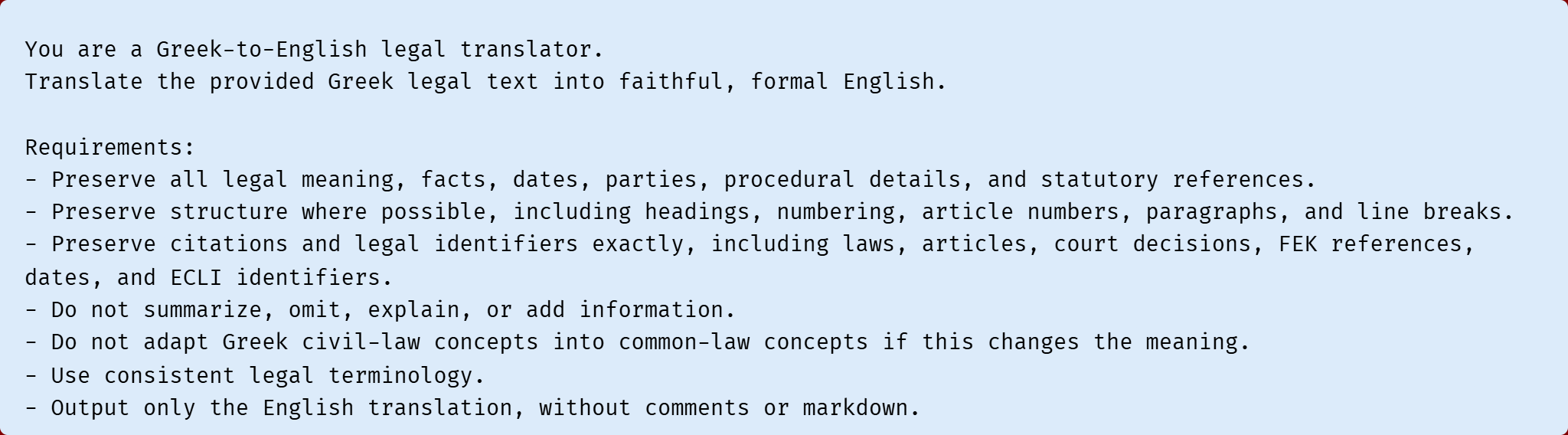}
    \caption{\textbf{Greek-to-English translation prompt}, applied to both queries and articles in the translation baseline.}
    \label{fig:en_translation_prompt}
\end{figure*}

\begin{figure*}[p]
    \centering
        \includegraphics[
        width=\textwidth,
        height=0.78\textheight,
        keepaspectratio
    ]{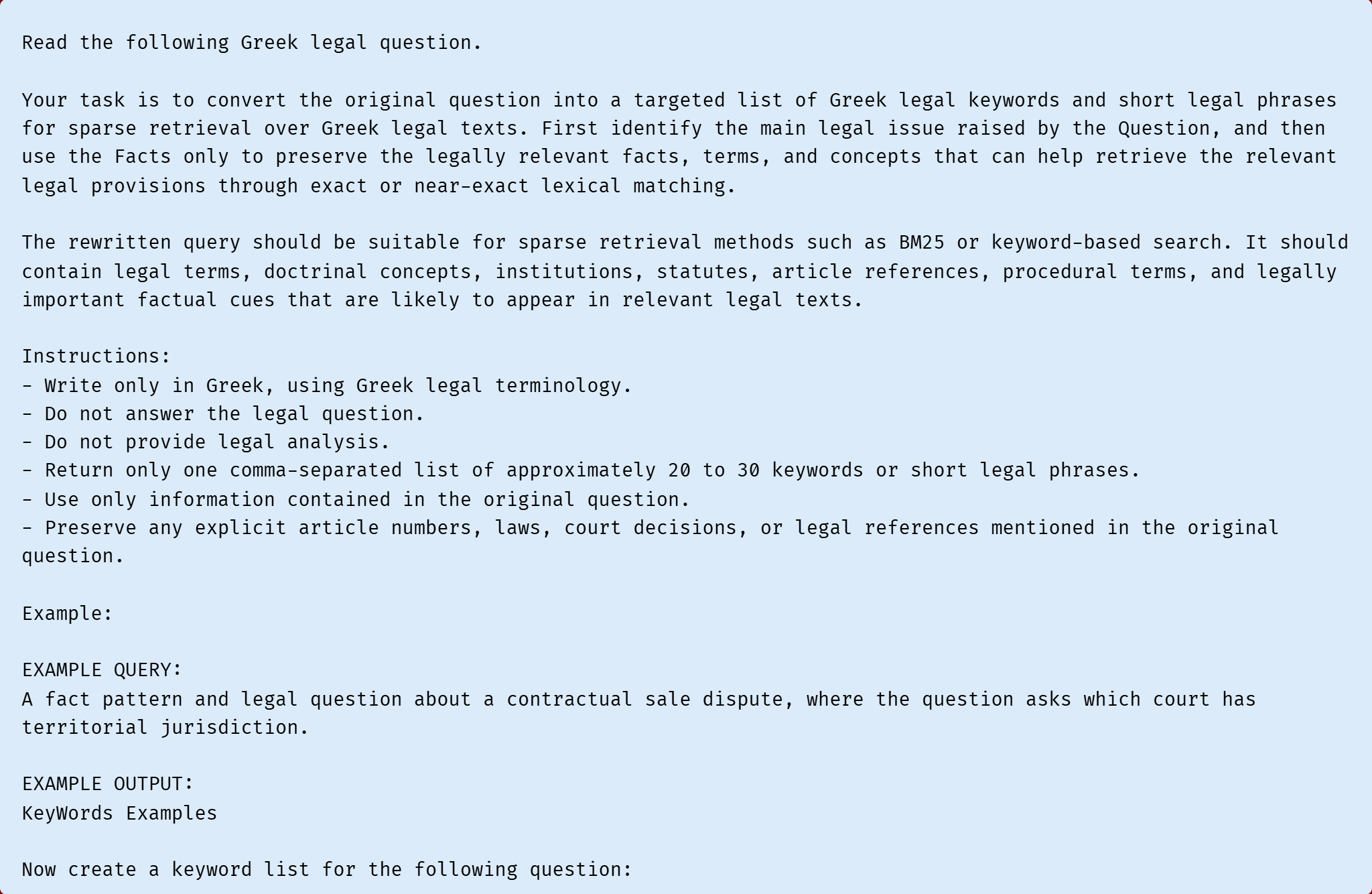}
    \caption{\textbf{Query reformulation prompt for sparse retrieval}. The output is a list of legal keywords and short phrases. Prompt translated from Greek to English for presentation purposes.}
    \label{fig:sparse_rewrite_prompt}
\end{figure*}

\begin{figure*}[p]
    \centering
        \includegraphics[
        width=\textwidth,
        height=0.78\textheight,
        keepaspectratio
    ]{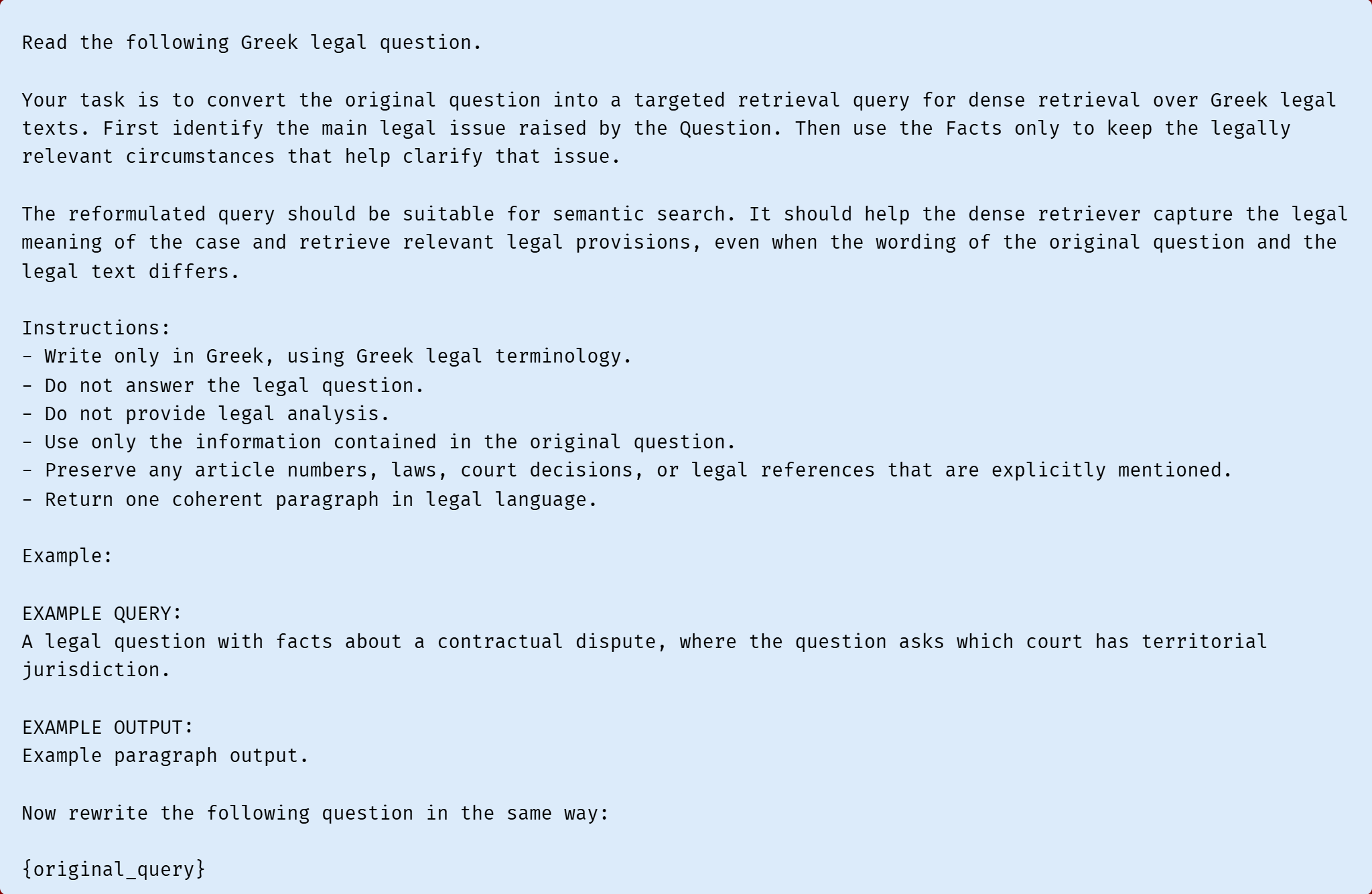}
    \caption{\textbf{Query reformulation prompt for dense retrieval}. The output is a prose statement of the legal issue. Prompt translated from Greek to English for presentation purposes.}
    \label{fig:dense_rewrite_prompt}
\end{figure*}

The prompt shown in Fig.~\ref{fig:en_translation_prompt} translates both queries and articles into English for the translation baseline. The prompts shown in Figures~\ref{fig:sparse_rewrite_prompt} and~\ref{fig:dense_rewrite_prompt} produce the reformulated queries. The two prompts differ in their output formats: the sparse prompt asks for legal terms and article headings, which \textsc{BM25} can match directly, while the dense prompt asks for a short prose statement of the legal issue.

\subsection{\textsc{ReAct-BM25}}

\begin{figure*}[p]
    \centering
     \includegraphics[
        width=\textwidth,
        height=0.78\textheight,
        keepaspectratio
    ]{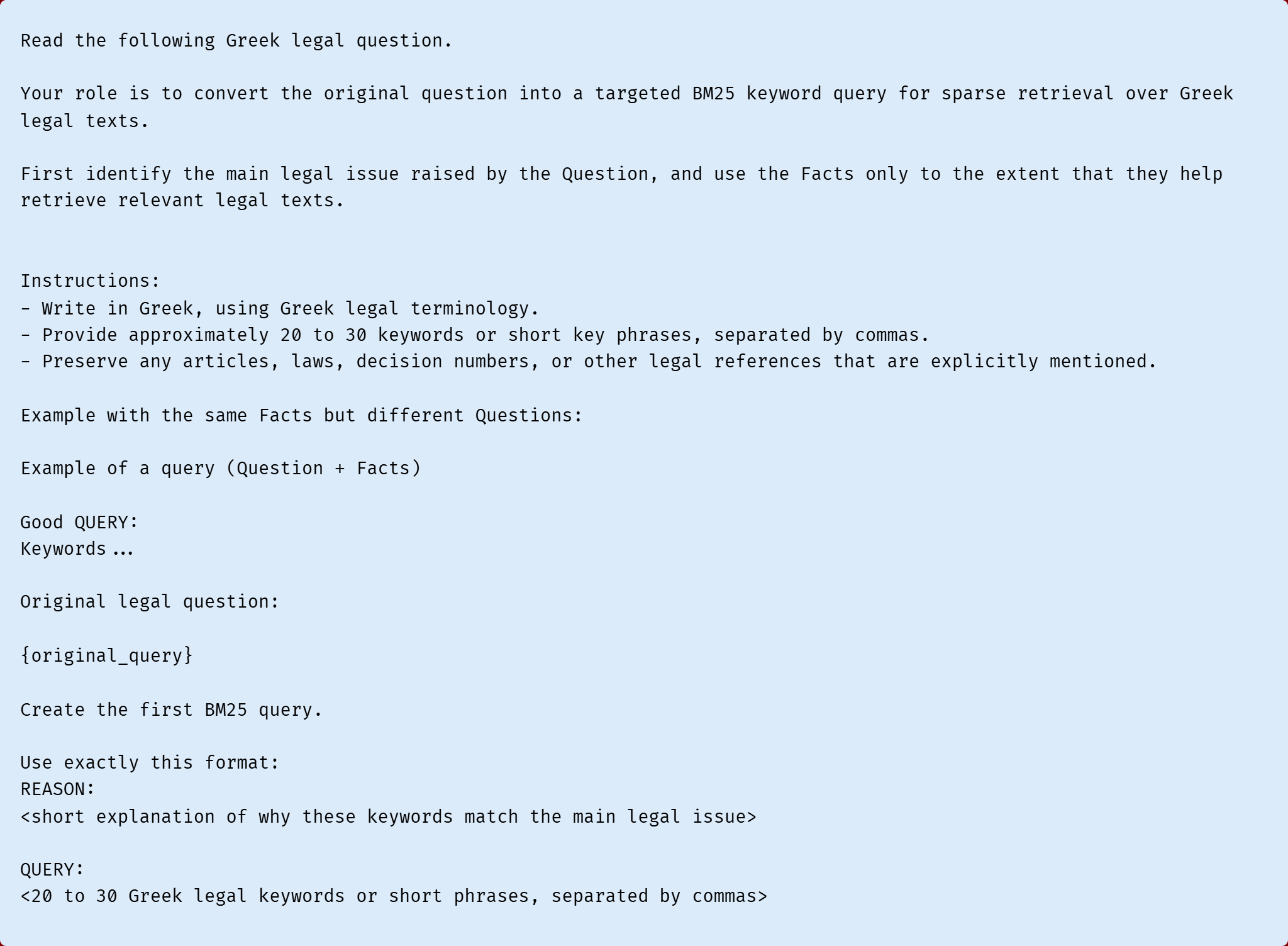}
    \caption{\textbf{Planner system prompt} for \textsc{ReAct-BM25}, fixed across all rounds. Prompt translated from Greek to English for presentation purposes.}
    \label{fig:react_planner_system_prompt}
\end{figure*}

\begin{figure*}[p]
    \centering
    \includegraphics[width=0.82\textwidth]{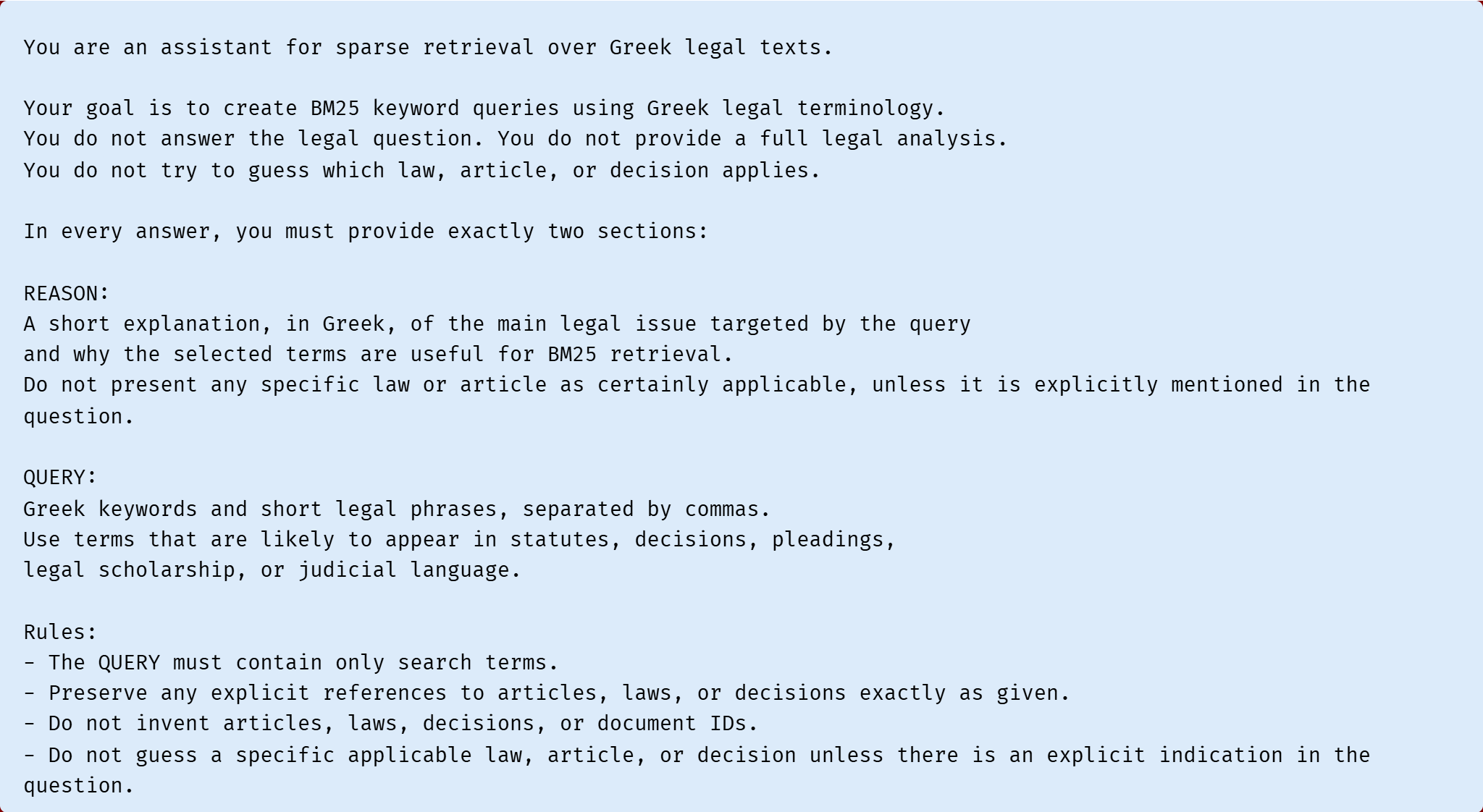}
    \caption{\textbf{First-round planner prompt} for \textsc{ReAct-BM25}. The planner sees the facts and produces an initial query. Prompt translated from Greek to English for presentation purposes.}
    \label{fig:react_planner_initial_prompt}
\end{figure*}

Four prompt templates define the planner-observer loop. The planner system prompt shown in Fig.~\ref{fig:react_planner_system_prompt} is a fixed instruction setting the planner's role and output format; it does not receive a query by itself. In round 1, this system prompt is paired with the first-round planner prompt shown in Fig.~\ref{fig:react_planner_initial_prompt}, which receives the original \texttt{Question + Facts} query and asks the planner to produce a Greek legal keyword query for \textsc{BM25-GreekStemmer}. \textsc{BM25-GreekStemmer} then searches the full retrieval pool and returns the top 100 candidate articles.

\begin{figure*}[p]
    \centering
        \includegraphics[
        width=\textwidth,
        height=0.78\textheight,
        keepaspectratio
    ]{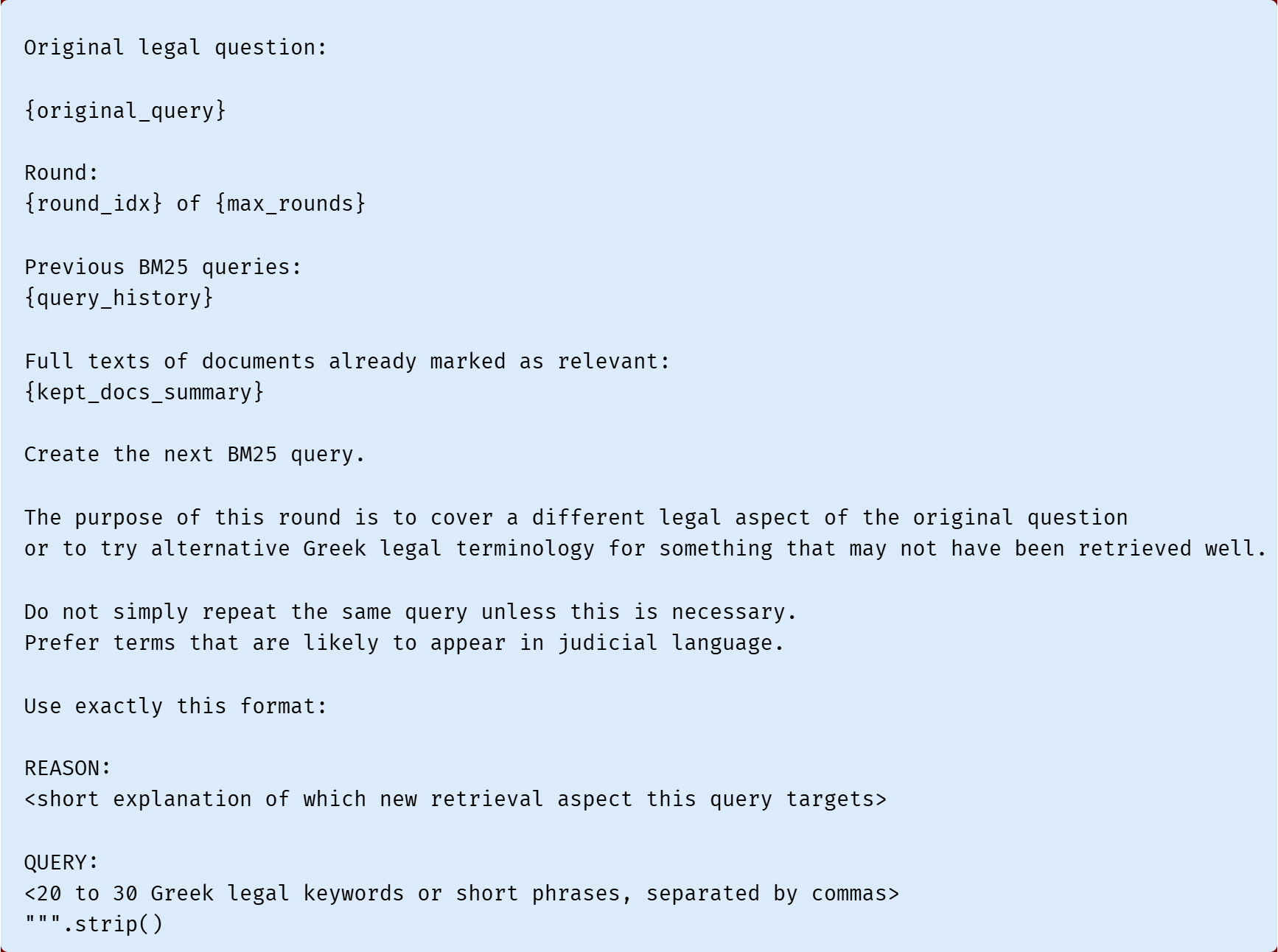}
    \caption{\textbf{Follow-up planner prompt} for \textsc{ReAct-BM25}, used in every round after the first. The planner also sees the articles kept so far. Prompt translated from Greek to English for presentation purposes.}
    \label{fig:react_planner_followup_prompt}
\end{figure*}

\begin{figure*}[p]
    \centering
        \includegraphics[
        width=\textwidth,
        height=0.78\textheight,
        keepaspectratio
    ]{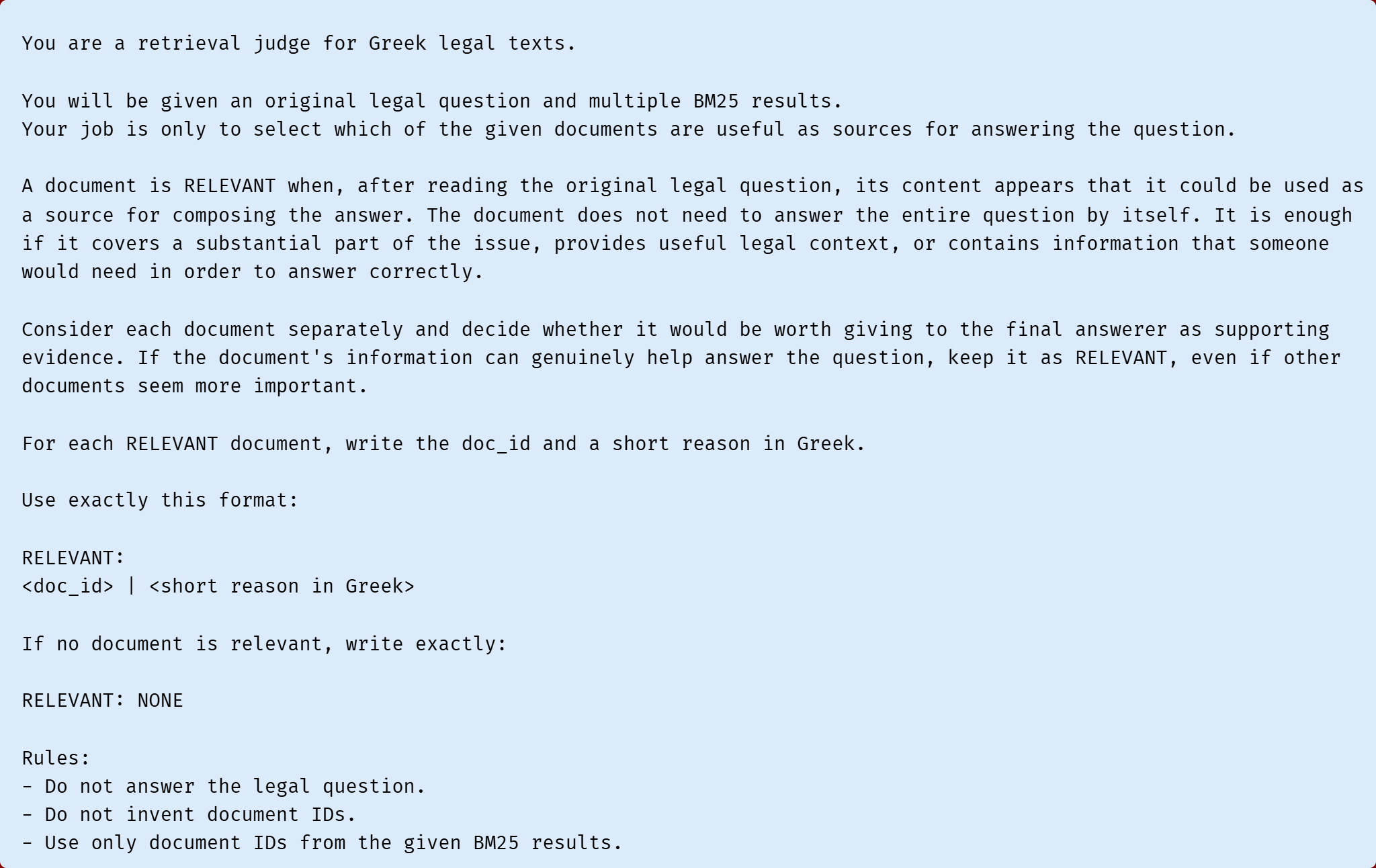}
    \caption{\textbf{Observer prompt} for \textsc{ReAct-BM25}. The observer selects which retrieved articles to keep and passes them to the next round. Prompt translated from Greek to English for presentation purposes.}
    \label{fig:react_observer_prompt}
\end{figure*}

The observer prompt shown in Fig.~\ref{fig:react_observer_prompt} receives the original \texttt{Question + Facts} query and the text of the candidate articles returned in that round, not the entire retrieval pool. It selects which candidates to retain as evidence. In later rounds, the planner uses the follow-up prompt shown in Fig.~\ref{fig:react_planner_followup_prompt}, which receives the original query, the previous planner-generated search queries, and the articles retained by the observer in earlier rounds. It then generates a new keyword query for the next \textsc{BM25-GreekStemmer} search. Thus information flows from planner to retriever, from retriever to observer, and from the observer's retained articles back to the planner in the next round. Because the observer prompt includes candidate article text, it accounts for most of the token cost reported in Table~\ref{tab:cost_analysis}.

\end{document}